\documentclass[12pt]{iopart}
\usepackage{iopams}
\usepackage{bm}
\usepackage{amsfonts}
\usepackage{amsthm}
\usepackage{mathtools}
\usepackage{enumerate} 
\usepackage{graphicx}
\usepackage{dcolumn}
\usepackage[utf8]{inputenc}
\usepackage{xcolor}
\usepackage{xspace} 
\usepackage{soul} 
\usepackage[normalem]{ulem} 
\usepackage{hyperref}
\hypersetup{
    colorlinks=true,
    linkcolor=blue,    
    urlcolor=cyan,
    citecolor=blue,
}

\newcommand{\Alfven}{{Alfv\'en}\xspace}
\newcommand{\iotabar}{\mbox{$\iota\!\!$-}}

\newcommand{\bB}{\mathbf{B}}

\newcommand{\gradpsi}{\nabla \psi}
\newcommand{\gradpsip}{\nabla \psi_p}
\newcommand{\gradt}{\nabla \vartheta}
\newcommand{\gradz}{\nabla \zeta}
\newcommand{\grada}{\nabla \alpha}

\newcommand{\gradc}{\nabla \chi}
\newcommand{\baralpha}{\bar{\alpha}}
\newcommand{\barpsi}{\bar{\psi}}
\newcommand{\bart}{\bar{\vartheta}}

\newcommand{\gradbaralpha}{\nabla \bar{\alpha}}
\newcommand{\gradbarpsi}{\nabla \bar{\psi}}
\newcommand{\gradbart}{\nabla \bar{\vartheta}}

\newcommand{\bdotgrad}{\bB \cdot \nabla }

\newcommand{\sn}{\text{sn}}
\newcommand{\am}{\text{am}}

\newcommand{\figref}[1]{{Figure \ref{#1}}}
\newcommand{\Figref}[1]{{Figure \ref{#1}}}
\newcommand{\secref}[1]{{Section \ref{#1}}}
\newcommand{\Eqref}[1]{{Equation \eqref{#1}}}

\newcommand{\modi}{\color{black}}

\newcommand{\norm}{\color{black}}

\definecolor{red}{rgb}{1,0,0}
\definecolor{gre}{rgb}{0,1,0}
\definecolor{blu}{rgb}{0,0,1}

\newcommand{\ZSQ}[1]{}

\begin{document}

\bibliographystyle{Science}

\title{The shear \Alfven continuum with a magnetic island chain in tokamak plasmas}

\author{Z. S. Qu$^{1}$
\footnote{Present Address: School of Physical and Mathematical Sciences, Nanyang Technological University, 637371 Singapore, Singapore.}
and M. J. Hole$^{1,2}$}

\address{$^1$Mathematical Sciences Institute, the Australian National University, Canberra ACT 2600, Australia}
\address{$^2$Australian Nuclear Science and Technology Organisation, Locked Bag 2001, Kirrawee DC NSW 2232, Australia}

\ead{zhisong.qu@ntu.edu.sg}




\vspace{10pt}

\begin{abstract}
The shear \Alfven continuum spectrum is studied for a tokamak with a single island chain using the ideal Magnetohydrodynamics (MHD) theory. We have taken into account the toroidal geometry and toroidal mode coupling with the island considered as a highly-shaped stellarator. Various new frequency gaps open up inside the island due to its asymmetry both poloidally and toroidally, such as the Mirror-induced \Alfven Eigenmode (MAE) gap and the Helicity-induced \Alfven Eigenmode (HAE) gap. We have shown that the MAE gap acts as the continuation of the outside Toroidal \Alfven Eigenmode (TAE) gap into the island. However, the combined TAE/MAE gap is getting narrower as the island grows, leaving only half of its original width with a moderate island size as much as 3.2\% of the minor radius. In addition, the two-dimensional eigenfunction of the continuum mode on the lower tip of the MAE gap now has highly localised structures around the island's long axis, 
contrary to the usual oscillatory global solutions found with no or a low level of toroidal asymmetry - an indication of the continuous spectrum becoming discrete and dense. These results have implications for the frequency, mode structure and continuum damping of global TAEs residing in the gap.
\end{abstract}

\section{Introduction}
Magnetically confined fusion plasmas contain significant fast populations originating from fusion products and external heating such as the neutral beam injection (NBI) and the ion cyclotron resonance heating (ICRH)~\cite{Fasoli2007}.
These fast particles, when slowed down, can excite a zoo of \Alfven eigenmodes as discrete solutions of the ideal Magnetohydrodynamics (MHD) spectrum in a process similar to the inverse Landau damping,
leading to enhanced fast ion transport and therefore worse energy output~\cite{Heidbrink2008}.
Of all the \Alfven eigenmodes, the most experimentally prolific is the Toroidicity-induced \Alfven Eigenmode (TAE)~\cite{Cheng1985, Cheng1986},
which resides in the band gaps of the shear \Alfven continuum spectrum induced by the poloidal modulation of the magnetic field and geometry.

The classic theory and numerical solvers of TAEs in tokamaks generally assume nested flux surfaces and perfect toroidal symmetry.
However, broken symmetry is introduced unavoidably, by the finite number of field coils or spontaneous instabilities such as tearing modes~\cite{Furth1973}, and deliberately, through the use of resonant magnetic perturbation (RMP) coils~\cite{Evans2006}, to suppress large explosive instabilities known as edge localised modes (ELMs)~\cite{Loarte2007}.
With the loss of symmetry and thus integrability, the field lines can tangle around a fixed-point, creating so-called magnetic islands, or when multiple islands overlap, regions of field line chaos.
The impact of symmetry-breaking fields on \Alfven eigenmodes is an emerging research topic. Several experiments in NSTX~\cite{Bortolon2013,Kramer2016} and KSTAR~\cite{Kim2020} have found an RMP field to either reduce or enhance the amplitude of the TAE, depending on RMP phasing and plasma conditions.
\modi
Existing works focused on the influence of RMP on the energetic particle distribution function~\cite{Garcia-Munoz2019} 
\norm
and the change of background plasma parameters such as the rotation, taking the mode frequency and structure to be the same as if the symmetry is not broken.  Nevertheless, islands and chaos could modify the frequency, mode structure and damping rate of the TAEs and thus affect energetic particle confinement. 

As an important first step, one needs to answer the question of how a single island chain changes the shear \Alfven continuum spectrum, in particular the TAE gap where global eigenmodes reside.
The continuum with a magnetic island has been studied in slab and cylindrical geometries~\cite{Biancalani2010prl,Biancalani2010,Biancalani2011,cook2015shear,Cook2015a,Yang2022}, 
\modi
where the change of the field strength poloidally,
as well as the coupling of modes with different toroidal numbers due to an absence of toroidal symmetry,
are neglected.
\norm
The idea is to separate the island chain from the rest of the plasma volume, construct a coordinate system within the island aligning with the magnetic surfaces inside, and then apply the same continuum equations as the outside.
In other words, one considers the island itself as a straight flux tube with its O point being the new magnetic axis with nested flux surfaces surrounding it.
The main finding is that the island has its own frequency gaps, with the dominant one being the Ellipticity-induced \Alfven Eigenmode (EAE) gap~\cite{Dewar1974}, or the Magnetic-island-induced \Alfven Eigenmode (MiAE) gap named by Biancalani \etal~\cite{Biancalani2010prl}, due to the elongation of the island.
Moreover, the lowest-frequency continuum accumulation point (CAP) is shifted up on the island separatrix, thanks to the strong poloidal mode coupling there.
A recent publication~\cite{Konies2022} extends the three-dimensional (3D) continuum code CONTI~\cite{Konies2012} to compute the continuum in Wendelstein 7-X with islands.
The discovery of the island EAE/MiAE gap prompts a further search for discrete MiAEs, with candidate modes being identified in TJ-II~\cite{Sun2015} 
\modi and J-TEXT~\cite{Liu2019} \norm
experimentally, and in Madison Symmetric Torus (MST)~\cite{Cook2016} using the SIESTA-\Alfven code~\cite{cook2015shear}, an extension of the SIESTA~\cite{Hirshman2011} 3D equilibrium code with a kinetic normalisation matrix.
\modi
Magnetic islands are also found to interact and excite Beta-induced \Alfven Eigenmodes (BAEs) in FTU~\cite{Buratti2005,Annibaldi2007} and later in HL-2A~\cite{Chen2011} and J-TEXT~\cite{Liu2019}.
\norm

Despite these great advancements, important questions regarding the connection between the inside/outside continuum remain unanswered.
For instance, does the TAE gap outside extends into the island chain? 
If the answer is yes, does the width of the inside gap match that of the outside?
One major limitation of the aforementioned analytical works is the absence of toroidicity and toroidal mode coupling.
In tokamak geometry, an island winds around the core of the plasma following the magnetic field lines, creating a complicated 3D magnetic structure with both poloidal and toroidal asymmetry.
It is no longer appropriate to be viewed as a straight tube, but rather, as a mini-stellarator.
One can therefore make use of the established methodology and knowledge of continuum in stellarators to compute and analyse that in the island chain.
In stellarators, the toroidal mode number is no longer a good quantum number, and, as a consequence, each eigenmode consists of multiple toroidal harmonics, while the toroidal coupling also gives rise to new gaps such as the Mirror-induced \Alfven Eigenmode (MAE) gaps~\cite{Nuhrenberg2000,Kolesnichenko2001} and the Helicity-induced \Alfven Eigenmode (HAE) gaps~\cite{Nakajima1992, Nuhrenberg1999}.
We will show in the paper that \modi within the ideal MHD theory \norm these new gaps are ultimately responsible for the continuation of the outside TAE gap into the island, 
whose width is determined by the interaction between them.
Finally, a high level of toroidal asymmetry could also change the continuous spectra into localised discrete ones~\cite{Salat1992}, 
which may have a fundamental impact on their interaction with existing global modes,
or create new ones.

In this work, we aim to study the shear \Alfven continuum in the presence of an island chain in tokamak geometry.
The current paper is organised as follows.
\secref{sec:theory} introduces the magnetic field, geometry and the equations for the shear \Alfven continuum.
\secref{sec:num} briefly describes the numerical scheme and benchmarks it against analytical results in the literature.
With the newly developed code, we compute the continuum both inside the island-stellarator and outside in the bulk of the plasma, as detailed in \secref{sec:results}.
The width of the combined gap and the eigenfunction of the mode are also investigated.
Finally, \secref{sec:conclusions} discusses the results and draws the conclusions.

\section{Theory}
\label{sec:theory}
\subsection{The magnetic field with an island chain}

We start with a magnetic field $\bB$ given by
\begin{equation}
 \bB = \gradpsi \times \gradt - \gradpsip \times \gradz,
 \label{eq:B}
\end{equation}
in which $\psi$ is the toroidal flux and will be used as the radial coordinate. 
The two angles $\vartheta$ and $\zeta$ are generalised angles in the poloidal and toroidal directions, respectively.
\modi
The physical quantities are by default in SI units.
The poloidal flux function $\psi_p$ is the superposition of an unperturbed axisymmetic equilibrium field and a surface-breaking perturbation and is given by
\begin{equation}
    \psi_p = \int_0^\psi \frac{d\psi'}{q(\psi')} + A \cos (m_0 \vartheta - n_0 \zeta),
\end{equation}
\norm
where $q$ is the safety factor, and $A$ is the amplitude of the flux perturbation,
while $m_0$ and $n_0$ label the helicity of the island chain.
\modi
We have used the ``constant-$\psi$'' approximation \cite{Rutherford1973} for a nonlinear tearing mode in which $A$ is assumed to be a constant. 
\norm
The contra-variant field components are given by
\begin{equation}
    J B^{\psi} = A m_0 \sin(m_0 \vartheta - n_0 \zeta), \quad J B^\vartheta = \frac{1}{q}, \quad JB^\zeta = 1,
    \label{eq:B_components}
\end{equation}
in which $J = (\gradpsi \times \gradt \cdot \gradz)^{-1}$ is the Jacobian of the coordinate system.

\begin{figure}[htbp]
    \centering
    \includegraphics[width=8cm]{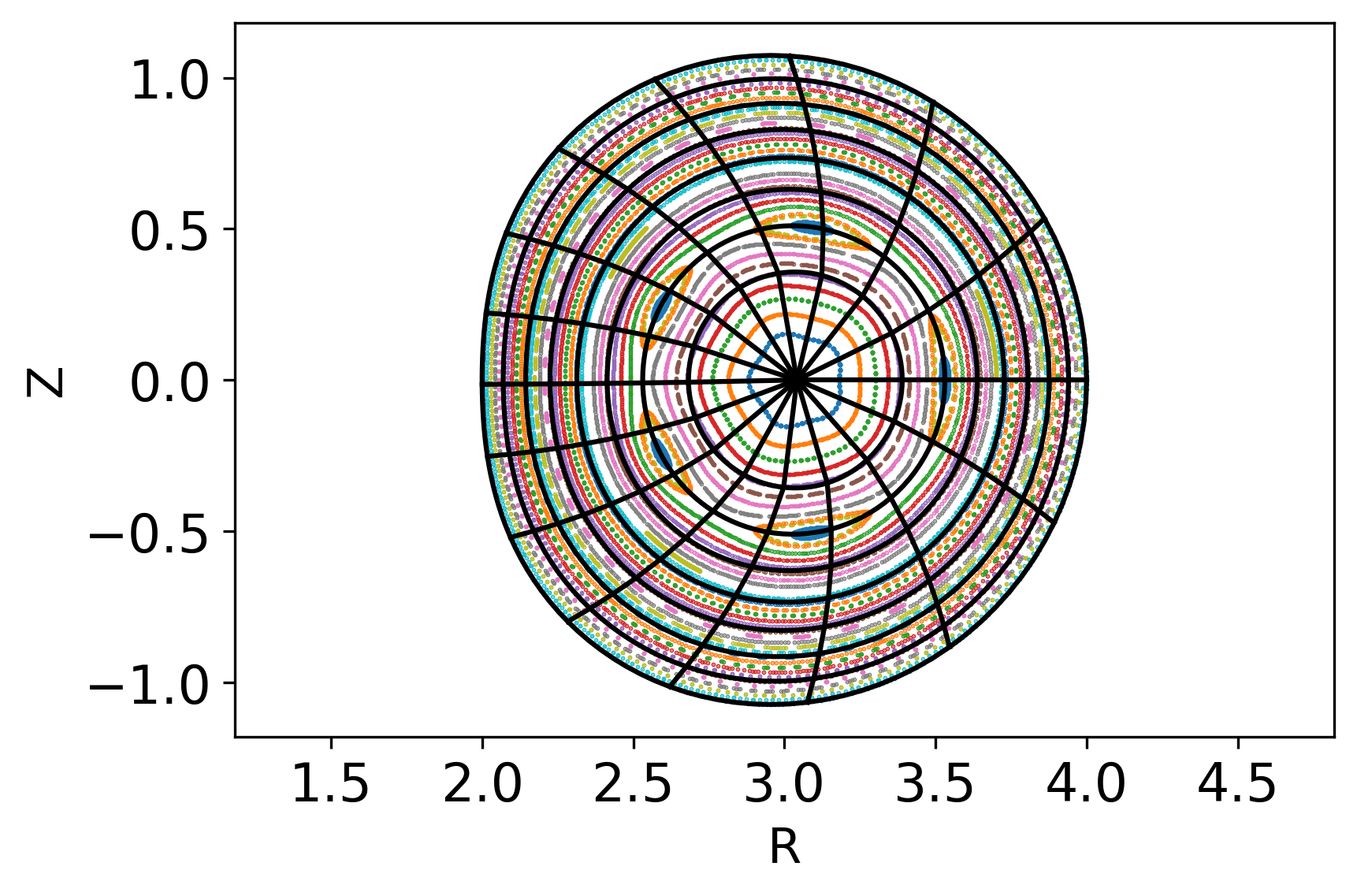}
    \caption{Poincar\'e plot of a $m_0=5, n_0=2, A=10^{-4}$ island chain with $dq/d\psi=4, \psi_0=0.125, a=1, R_0=3$, $B_0 = 1$ \modi and a linear rotational transform profile. Constant $\psi$ and $\vartheta$ surfaces are indicated by black solid lines.\norm}
    \label{fig:poincare}
\end{figure}

\modi
The unperturbed geometry is a large aspect ratio, circular-cross-section tokamak.
The cylindrical coordinates $(R, \varphi, Z)$ are written in terms of the toroidal coordinates $(\psi, \vartheta, \zeta)$ with the relationship given by~\cite{Berk1992}
\begin{eqnarray}
    R = R_0 + r \cos \vartheta  - \Delta(r) + r \eta(r) (\cos 2 \vartheta - 1), \label{eq:R}\\
    \varphi = - \zeta, \\
    Z = r \sin \vartheta + r \eta(r) \sin 2 \vartheta \label{eq:Z},
\end{eqnarray}
where $R_0$ is the major radius of the plasma boundary,
with the metric tensor and the Jacobian given in \ref{app:metrics}.
The boundary of the plasma is circular with a radius of $a$.
The unperturbed toroidal flux is related to the radius $r$ by
\begin{equation}
    \psi = \frac{B_0}{2} r^2,
\end{equation}
where $B_0$ is the field strength on axis.
The flux surfaces are approximated by circles with radius $r$ 
and their centres are shifted from the centre of the boundary by a distance $\Delta(r)$.
This shift, known as the Shafranov shift, 
is a consequence of a non-zero pressure gradient and current density,
and is determined by solving the Grad-Shafranov equation~\cite{goedbloed2010advanced}.
Its derivative with respect to $r$ in the zero pressure limit is given by
\begin{equation}
    \Delta'(r) = \frac{q^2}{R_0 r^3} \int_0^{r} \frac{r^3}{q^2} dr,
\end{equation}
and with a zero or moderate shear $\Delta' \approx r/(4R_0)$.
The quantity $\eta(r) = (r/R_0 + \Delta')/ 2$ and its existence in \eqref{eq:R} and \eqref{eq:Z} is to ensure $\vartheta$ and $\zeta$ are straight-field-line angles to order $O(\epsilon)$.
This makes $\vartheta$ slightly different from the geometric poloidal angle.
A demonstration of the constant $\psi$ and $\vartheta$ surfaces is given by the black lines in \figref{fig:poincare}.
We note that the generalised toroidal angle $\zeta$ coincides with the (negative) true toroidal angle $\varphi$ and therefore the straight-field-line coordinates in this paper are PEST coordinates.

If $A=0$, then $B^\psi=0$ and $\psi_p$ is a function of $\psi$ only, meaning that the \modi unperturbed \norm magnetic field in \eqref{eq:B} is completely integrable, i.e. the field lines are lying on concentric, nested surfaces known as the flux surfaces labelled by $\psi$.
Moreover, $\vartheta$ and $\zeta$ are straight-field-line angles,
such that $B^\zeta/B^\vartheta = q(\psi)$ is a constant on each flux surface. 
When $|A| > 0$, the coordinate system is kept as it is,
while a magnetic field perpendicular to the constant $\psi$ surfaces is introduced.
\norm
An island chain will develop around the flux surface where 
\begin{equation}
    q(\psi_0) = q_0 = \frac{m_0}{n_0},
\end{equation}
in which $\psi_0$ is the radial location of resonance in terms of the unperturbed radial coordinate.
The integer $m_0$ gives the number of O points/X points on a toroidal cross-section, while $n_0$ gives the field period in the toroidal direction.
A Poincar\'e plot of an $m_0=5, n_0=2, A=10^{-4}$ island chain with $\psi_0=0.125, a=1, R_0=3$, $B_0 = 1$, $dq/d\psi=4$ and a linear rotational transform profile is overplotted in \figref{fig:poincare}.
\modi
The Poincar\'e plot is constructed by field-line tracing,
i.e. solving the ordinary differential equation $d\mathbf{X}/dl = \bB(\mathbf{X})$ for a number of initial locations,
where $\mathbf{X}$ is the location of a point on a field line and $l$ is a distance-like variable along the field line, and recording a point whenever a field line penetrate the $\varphi=0$ cross section.
Now with the island chain neither $\psi$ nor $\psi_p$ is a good flux label,
as the flux surfaces shown by the Poincar\'e plot are no longer aligned with the coordinate surfaces.
Also, the angles $\vartheta$ and $\zeta$ are not straight-field-line angles any more.
The next step is to construct a new radial coordinate aligning with the flux surfaces and new straight-field-line angles both inside and outside the island.
\norm
\modi

\norm

\subsection{Straight-field-line coordinates}
\label{sec:SFL_coordinates}

Now let $\alpha = \vartheta - \zeta/q_0$ be the helical angle labelling the rotation around the axis of the island. One can rewrite \eqref{eq:B} as
\begin{equation}
 \bB = \gradpsi \times \grada - \gradc \times \gradz,
\end{equation}
where the helical flux $\chi$ is given by
\begin{equation}
    \chi( \alpha, \psi) = \psi_p - \psi / q_0 - \chi_0 = \int_{\psi_0} ^\psi \left(\frac{1}{q} - \frac{1}{q_0} \right) d\psi' + A \cos (m_0 \alpha),
\label{eq:chi}
\end{equation}
with $\chi_0$ being an integration constant.
If the rotational transform $\iotabar = 1/q$ is a linear function of $\psi$,
\eqref{eq:chi} has a simpler form given by
\begin{equation}
    \chi( \alpha, \psi) = - \frac{q'}{2q_0^2}(\psi - \psi_0)^2 + A \cos(m_0 \alpha).
\label{eq:chi_quadratic}
\end{equation}
We shall take this assumption for the rest of the paper.

The equations of the magnetic field lines are identical to the canonical equations of a Hamiltonian system with the Hamiltonian given by
\begin{equation}
    H(\alpha, \psi) = \chi( \alpha, \psi),
\end{equation}
where $\alpha$ is the position, $\psi$ is the corresponding momentum, and $\zeta$ is now the pseudo-time variable.
Finding the straight-field-line coordinates is equivalent to constructing action-angle coordinates for the Hamiltonian system,
such that the Hamiltonian is a function of the action $\barpsi$ independent of the corresponding angle $\baralpha$,
i.e. $H(\alpha, \psi) = H(\barpsi)$.
The canonical equations are therefore
\begin{equation}
    \frac{d \baralpha}{d \zeta} = \frac{B^{\baralpha}}{B^\zeta}=\left.\frac{\partial H}{\partial \barpsi}\right|_{\baralpha} = \Omega(\barpsi), \quad
     \frac{d \barpsi}{d \zeta} = \frac{B^{\barpsi}}{B^\zeta}= -\left.\frac{\partial H}{\partial \baralpha}\right|_{\barpsi} = 0,
\end{equation}
such that $\barpsi$ is a flux label and $\baralpha$ is a straight-field-line angle.
The frequency $\Omega$ is the rotational transform with respect to the axis (O point) of the island.
The physical meaning of the action $\barpsi$ is the toroidal flux enclosed by the new flux surfaces,
while $\baralpha$ represents the phase on an island.
For example, on the separatrix $\baralpha = \pi /2$ and $3 \pi /2$ always correspond to the two X points regardless of the toroidal angle.

Assuming $A > 0$ and $q'>0$ for simplicity,
one can define $\kappa$ as 
\begin{equation}
    \kappa =\frac{-\chi + A} {2A},
\end{equation}
such that $\kappa > 1$ outside the island and $\kappa < 1$ inside,
while $\kappa=1$ corresponds to the separatrix containing the X point.
The full width of the island in terms of $\psi$ can be obtained by substituting $\kappa=1$ into \eqref{eq:chi_quadratic} and taking the difference between the maximum and the minimum value of $\psi$ for a fixed $\chi$, written as
\begin{equation}
    w = 4 \sqrt{\frac{A q_0^2}{q'}}.
\end{equation}
With the form of $\chi$ given in \eqref{eq:chi_quadratic},
$H$ is equivalent to the Hamiltonian of a pendulum,
whose action-angle coordinates are well-known.
Please refer to \ref{app:island_action_angle} for details.
We note that the approach here is similar to~\cite{Cook2015a}
and references therein~\cite{Hegna1992, Hegna2011}

Outside the island, the new flux surfaces enclose the original magnetic axis,
while they only enclose the centre of the island inside the island chain.
The value of $\barpsi$ is given in terms of $\kappa$ by
\begin{equation}
    \barpsi = 
    \left\{
    \begin{array}{ll}
         \pm \frac{w}{\pi} \sqrt{\kappa} E(\kappa^{-1}) + \psi_0, & \kappa > 1   \\
         \frac{2 w}{m_0 \pi}   [(\kappa - 1) K(\kappa) + E(\kappa)], & \kappa \le 1
    \end{array}
    \right.
    \label{eq:toroidal_flux}
\end{equation}
where $K(x)$ and $E(x)$ are the complete elliptic integral of the first and second kind, respectively.
The surfaces between the island chain and the magnetic axis take the negative sign and the ones beyond the island take the positive sign.
In the limit $\kappa \rightarrow 1^+$, $\barpsi \rightarrow \pm w /\pi + \psi_0$. This implies a discontinuity of $\Delta \barpsi = 2w/\pi$ across the island chain,
corresponding to the toroidal flux occupied by the island chain.
Similarly, in the limit $\kappa \rightarrow 1^-$, $\barpsi \rightarrow \barpsi_{\text{isl}} = 2w /(\pi m_0)$.
The extra factor of $1/m_0$ is coming from the fact that there are $m_0$ islands in the island chain on the same toroidal plane, so each island will occupy $1/m_0$ of the total toroidal flux of all the islands on that cross section.
The local rotational transform can be computed by taking the derivative of \eqref{eq:toroidal_flux}, given by
\begin{equation}
    \Omega = 
    \left\{
    \begin{array}{ll}
         \mp \Omega_0 \frac{\pi \sqrt{\kappa}}{K(\kappa^{-1})}, & \kappa > 1   \\
         - \Omega_0 \frac{m_0 \pi}{2 K(\kappa)}, & \kappa \le 1
    \end{array}
    \right.
\end{equation}
where $\Omega_0 = \sqrt{\frac{A q' }{q_0^2}}$.



After the action-angle transformation, 
the magnetic field can be written in the new straight-field-line coordinates as
\modi
\begin{equation}
    \bB = \gradbarpsi \times \gradbaralpha - \Omega(\barpsi) \gradbarpsi \times \gradz.
\end{equation}
\norm
Outside the island chain, one can define $\bart = \baralpha + \zeta / q_0$,  such that 
\begin{equation}
    \bB = \gradbarpsi \times \gradbart - [\Omega(\barpsi) + q_0^{-1}]  \gradbarpsi   \times \gradz.
\end{equation}
and
\begin{equation}
        J B^{\barpsi} =0, \quad J B^{\bart} = q_0^{-1} + \Omega(\barpsi), \quad J B^\zeta = 1.
\end{equation}
Note that $\baralpha$ is a helical angle so that the origin $\baralpha = 0$ is winding helically along with the island with respect to the magnetic axis as $\zeta$ increases, 
while $\bart$ is a regular poloidal angle and its origin stays almost in the same place on different toroidal planes.
The effective $q$ profile is $q(\barpsi) = [\Omega(\barpsi) + q_0^{-1}]^{-1}$.

Inside the island, we will use $\baralpha$ instead of $\bart$.
\modi
The (helical) rotational transform or inverse $q$ profile is therefore $\Omega$.
\norm
However, there are $m_0$ islands on the same toroidal cross-section for the same $\zeta$ so the coordinate is not single-valued.
This problem can be resolved by extending the period of $\zeta$ from $2 \pi$ to $2 m_0 \pi$, consistent with the toroidal length of the island before it closes back to itself.
This will only affect the boundary condition of the mode later on in the calculation.




\subsection{Shear \Alfven continuum}
\label{sec:theory_continuum}
In a large aspect ratio, zero-beta tokamak plasma with a constant current profile, the linear MHD equations for shear \Alfven waves can be simplified into~\cite{Berk1992, Rosenbluth1975}
\begin{eqnarray}
    -\omega^2 \nabla \cdot \left( \frac{\mu_0 \rho}{B^2} \nabla_\perp \Phi \right) 
&-& B \nabla_\parallel \frac{1}{B^2} \nabla \cdot \left( B^2 \nabla_\perp \frac{\nabla_\parallel \Phi}{B} \right) \nonumber\\
&-& \left[\bB \times \nabla \left(\frac{1}{B} \nabla_\parallel \Phi \right) \right] \cdot \nabla \left( \frac{\mu_0 \mathbf{J} \cdot \bB}{B^2} \right)
=0,
\label{eq:full_wave}
\end{eqnarray}
where $\omega$ is the wave frequency, $\Phi$ the perturbed electrostatic potential,
$B = |\bB|$ the field strength, $\mathbf{J} = \mu_0^{-1} \nabla \times \bB$ the equilibrium current, $\rho$ the plasma mass density,
$\mu_0$ the vacuum permeability constant,
\modi
$\nabla_\parallel = \mathbf{b} \cdot \nabla$ the parallel gradient operator, $\mathbf{b} = \bB / B$ the unit vector parallel to $\bB$ and $\nabla_\perp = \nabla - \mathbf{b} \nabla_\parallel$ the perpendicular gradient operator.
\norm
To construct \eqref{eq:full_wave},
one first eliminates the fast magneto-sonic waves by writing 
the perturbed electric field into $\tilde{\mathbf{E}} = -\nabla \Phi + \partial_t \tilde{A}_\parallel$, in which $\tilde{A}_\parallel$ is the parallel component of the perturbed vector potential and $\partial_t$ the time derivative.
Next, noting that $\tilde{\mathbf{E}} \cdot \bB=0$ in the ideal Ohm's law, $\tilde{A}_\parallel$ can be written in terms of $\Phi$.
Using the ideal Ohm's law, the fluid velocity can be replaced by $\Phi$, which is then substituted into the momentum equation to compute the perturbed perpendicular current $\tilde{\mathbf{J}}_\perp$.
The perturbed parallel current $\tilde{J}_\parallel$ is obtained by taking the curl of the vector potential $\tilde{\mathbf{A}} = \tilde{A}_\parallel \mathbf{b}$ directly.
Finally, the perpendicular and parallel currents are put together in the quasi-neutrality condition $\nabla \cdot \tilde{\mathbf{J}} = 0$ to produce \eqref{eq:full_wave}.
We have ignored the effect of a non-zero beta such as the up-shift of the lowest-frequency continuum accumulation point (CAP) due to the coupling between the shear \Alfven waves and the sound waves through the geodesic curvature. 

\Eqref{eq:full_wave} is solved as an eigenvalue problem,
i.e. we look for eigenvalues $\omega$ and eigenfunctions $\Phi$ that satisfy the ideal wall boundary condition $\Phi = 0$ at the plasma edge.
When $\omega$ belongs to the discrete spectrum, for example a toroidal \Alfven eigenmode,
$\Phi$ is a smooth global solution across the whole plasma region.
When $\omega$ belongs to the continuum spectrum, $\Phi$ has a quasi-mode solution singular across one or multiple flux surfaces.
We plot the frequency as a function of the radial coordinate on which the mode is singular, known as the \Alfven continuum.

To compute the \Alfven continuum, one notes that in \eqref{eq:full_wave} the highest derivative with respect to the flux surfaces will dominate all other terms due to the existence of the singularity.
In our coordinate system derived in \secref{sec:SFL_coordinates},
\eqref{eq:full_wave} can be reduced to
\begin{equation}
    \omega^2 \rho \frac{|\gradbarpsi|^2}{B^2} \Phi + \bdotgrad \left( \frac{|\gradbarpsi|^2}{B^2} \bdotgrad \Phi \right) = 0,
    \label{eq:continuum}
\end{equation}
in which
\begin{equation}
    \bdotgrad=  \frac{1}{J} (q^{-1} \partial_{\bart} + \partial_\zeta) = \frac{1}{J} (\Omega \partial_{\baralpha} + \partial_\zeta).
    \label{eq:bdotgrad}
\end{equation}
Here, we take the plasma density $\rho$ to be a constant throughout the plasma volume and note that
\begin{eqnarray}
    \Omega^2 |\gradbarpsi|^2 &=& |\nabla \chi|^2 \nonumber\\
    &=& \frac{q_0'^2}{q_0^4}(\psi-\psi_0)^2 |\gradpsi|^2 + A^2m_0^2 \sin^2 m_0 \alpha 
    \left(|\gradt|^2 + \frac{|\gradz|^2}{q_0^2} \right) \nonumber\\
    &+& 2 \frac{q_0'}{q^2_0}(\psi - \psi_0) A m_0 \sin m_0 \alpha (\gradt \cdot \gradpsi),
    \label{eq:gradpsi2}
\end{eqnarray}
which can be computed using \eqref{eq:metric_1}-\eqref{eq:metric_4}.
The Jacobian $J$ in \eqref{eq:bdotgrad} is identical to the form in \eqref{eq:B_components} and \eqref{eq:metric_5} noting that $\gradbarpsi \times \gradbaralpha = \gradpsi \times \grada$.

To solve \eqref{eq:continuum}, one also needs the boundary condition in $\bart$ (or $\baralpha$) and $\zeta$.
Outside the island chain, $\bart$ and $\zeta$ are poloidal and toroidal angles written in PEST coordinate, respectively and therefore,
\begin{equation}
    \Phi(\bart) = \Phi(\bart + 2 \pi), \quad
    \Phi(\zeta) = \Phi(\zeta + 2 \pi).
\end{equation}
We can expand $\Phi$ into its Fourier harmonics given by
\begin{equation}
    \Phi(\bart, \zeta) = \sum_{m} \sum_{n} \Phi_{m,n} e^{i m \bart - i n \zeta},
    \label{eq:fourier1}
\end{equation}
where $m$ and $n$ are integers.
Inside the island chain,
we note that the period of the toroidal angle has been expanded to $2m_0 \pi$, which leads to the boundary conditions given by
\begin{equation}
    \Phi(\baralpha) = \Phi(\baralpha + 2 \pi), \quad
    \Phi(\zeta) = \Phi(\zeta + 2 m_0 \pi).
\end{equation}
Thus, $\Phi$ can instead be expanded into
\begin{equation}
    \Phi(\baralpha, \zeta) = \sum_{m} \sum_{n} \Phi_{m,n} e^{i m \baralpha - i \frac{n}{m_0} \zeta}.
    \label{eq:fourier2}
\end{equation}

\section{Numerical solutions and benchmark}
\label{sec:num}

\subsection{Numerical implementation}
Substituting the Fourier expansions \eqref{eq:fourier1} and \eqref{eq:fourier2} into \eqref{eq:continuum}, multiplying both sides by $J$ and the complex conjugate of each Fourier basis, and integrating it over the two angles, one converts \eqref{eq:continuum} into a generalised matrix eigenvalue problem.
The eigenvalues and eigenvectors are then computed by the QR algorithm embedded in the SciPy package. 
In the actual computations, we have only included a selected range of $m$ and $n$ such that adding more harmonics will not affect the results in the frequency range of interest.
A typical run with 100 poloidal modes and 10 toroidal modes on 100 flux surfaces takes less than 2 minutes on a laptop.

\modi
We note that the numerical method we have implemented here is similar to COBRA~\cite{Kolesnichenko2001} and STELLGAP~\cite{Spong2003},
but it has the advantage of being able to compute the continuum both inside and outside the island chain conveniently at the same time.
One could in principle use standard stellarator continuum codes for inside and/or outside the island separately if one can construct an equilibrium for either region in VMEC~\cite{Hirshman1983} or similar 3D equilibrium codes.
Recently, a module similar to this paper was independently developed~\cite{Konies2022} for the CONTI code to compute the continuum for inside/outside the island chain given the island size and an unperturbed equilibrium field.
\norm

\subsection{Benchmark}
In the cylindrical limit without pressure, $m$ and $n$ are good quantum numbers. The frequencies of the continuum are given as a function of $q$ by
\begin{equation}
    \frac{\omega}{\omega_{A0}} = \left| \frac{m}{q} - n \right|,
\end{equation}
in which $\omega_{A0} = B_0/(\sqrt{\mu_0 \rho} R_0)$ is the \Alfven angular frequency on the magnetic axis.
This would mean that modes with $m=Nm_0$ and $n=Nn_0$ for any given integer $N$ will have $\omega = \omega_{\text{CAP}} = 0$ on flux surfaces where $q = q_0 = m_0/n_0$.
The frequency $\omega_{\text{CAP}}$ is known as the low-frequency continuum accumulation point (CAP).
Biancalani \etal~\cite{Biancalani2011} found that on the island separatrix where $q = q_0$, 
$\omega_{\text{CAP}}$ is shifted up from zero to
\begin{equation}
    \frac{\omega_{\text{CAP}}}{\omega_{A0}} = \frac{n_0 q' w}{4 q_0} = n_0 \sqrt{A q'},
    \label{eq:w_CAP}
\end{equation}
when converted into our coordinates and notations.
The same up-shift is later confirmed by Cook and Hegna~\cite{Cook2015a}. 

\begin{figure}[htbp]
    \centering
    \includegraphics[width=8cm]{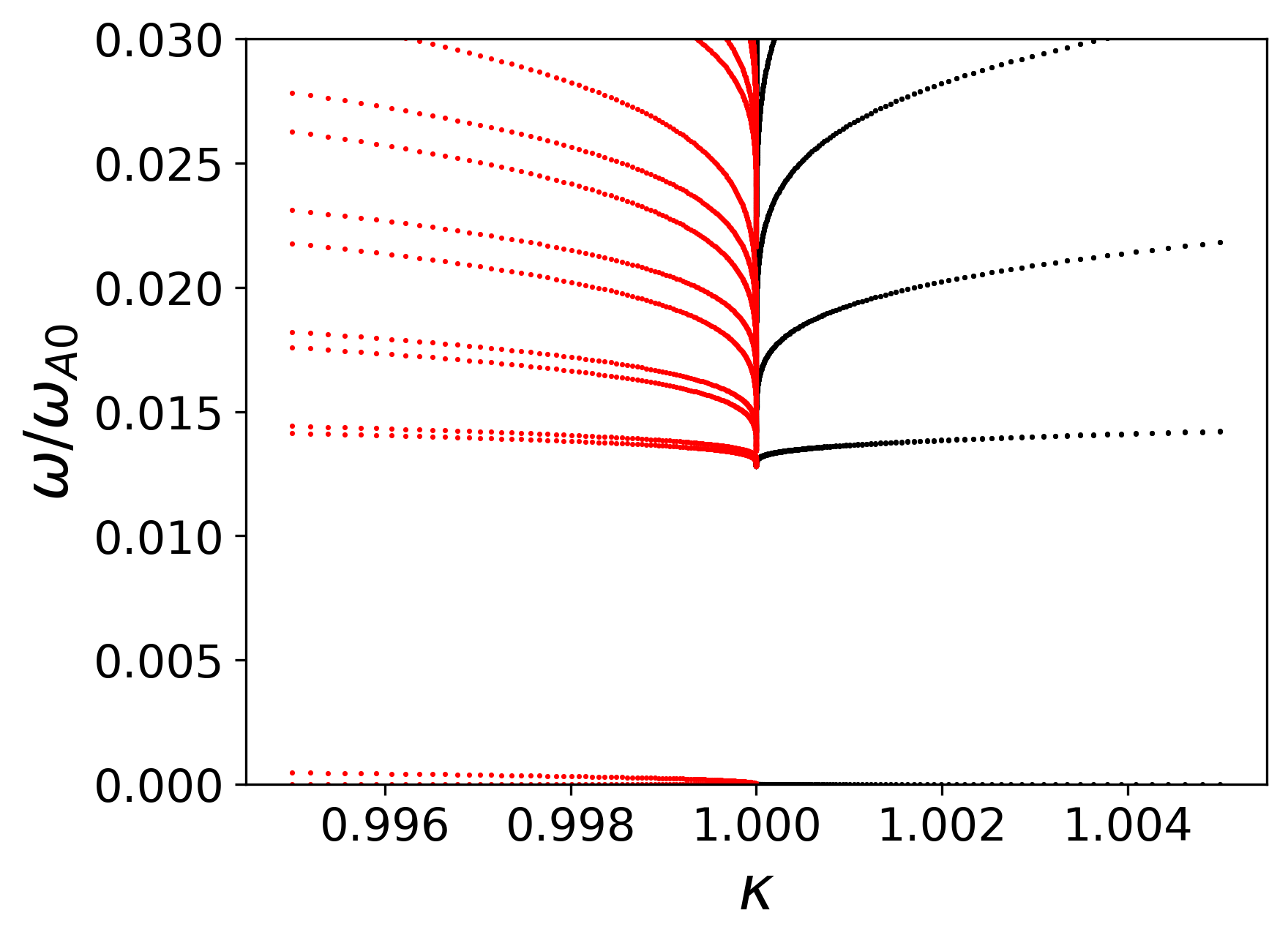}
    \caption{The shear \Alfven continuum near the separatrix 
    for $q'=4, \psi_0=0.125, a=1, R_0=1000, B_0 = 1, m_0=5, n_0=2, q_0=5/2$ and $A=1\times 10^{-5}$. }
    \label{fig:CAP}
\end{figure}

In this section, we will benchmark our code with the results of Biancalani \etal. 
To approach the limit of slab geometry assumed by the aforementioned works, we choose a very large aspect ratio $a/R_0=0.001$ and a small island width $w < 0.05 \psi_{\text{edge}}$, where $\psi_{\text{edge}}$ is the toroidal flux on the boundary of the plasma.
In such a limit, the effect of toroidicity can be ignored,
while the island chain and its neighbourhood can be locally approximated by a slab.
We have computed the continuum for $q'=4, \psi_0=0.125, a=1, R_0=1000$ and $B_0 = 1$.
In \figref{fig:CAP} we have plotted the continuum frequency in the region $0.995 \le \kappa \le 1$ (just inside the island) and $1 < \kappa \le 1.005$ (just outside the island) for $m_0=5, n_0=2, q_0=5/2$ and $A=10^{-5}$ ($w/\psi_{\text{edge}} = 0.032$).
For outside the island chain, the chosen Fourier harmonics are $m=-7m_0$ to $m=7m_0$ with an increment of $m_0$, and $n=-7n_0$ to $7n_0$ with an increment of $n_0$.
For inside, we compute the continuum with $m=-7$ to $7$ and $n=0$.
Same as Biancalani \etal, we found that all the continuum branches converge to an island-modified, non-zero $\omega_{\text{CAP}} \approx 0.013 \omega_{A0}$ on the separatrix, except the lowest frequency branch.

\begin{figure}[htbp]
    \centering
    \includegraphics[width=8cm]{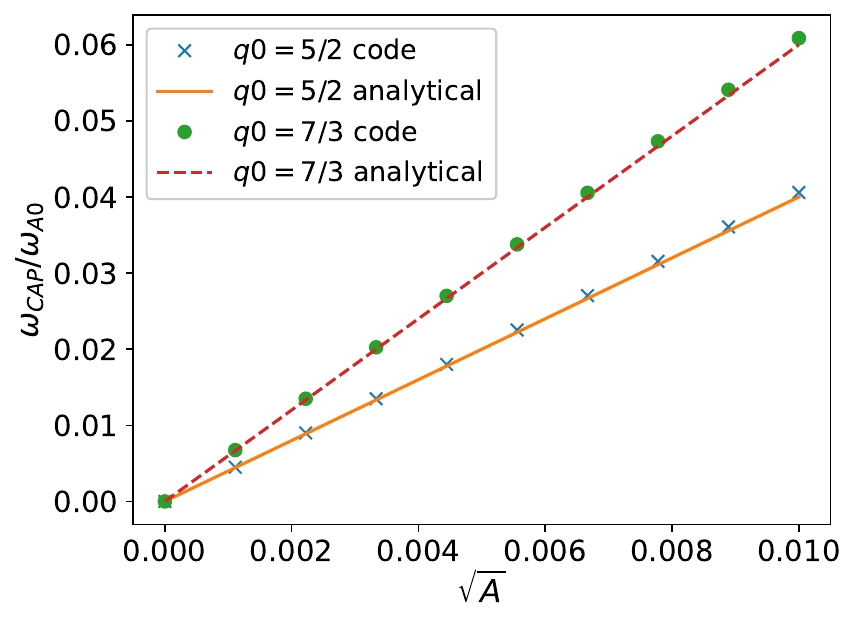}
    \caption{The comparison of the CAP frequency $\omega_{\text{CAP}}$ as a function of $\sqrt{A}$ (proportional to the island size) between the code and analytical results in \eqref{eq:w_CAP},
    for two different island helicities: $m_0=5, n_0=2, q_0=5/2$ and $m_0=7, n_0=3, q_0=7/3$. 
    Other parameters are the same as \figref{fig:CAP}.}
    \label{fig:benchmark}
\end{figure}

We have also scanned $\omega_{\text{CAP}}$ as a function of $\sqrt{A}$ for $0 \le A \le 10^{-4}$ for two different island helicities: $m_0=5, n_0=2, q_0=5/2$ and $m_0=7, n_0=3, q_0=7/3$ as demonstrated in \figref{fig:benchmark}.
It is evident that $\omega_{\text{CAP}}$ scales linearly with $\sqrt{A}$, which is proportional to the island width.
These results are compared to the analytical solution given in \eqref{eq:w_CAP}, showing a very good agreement.

\section{Continuum in the presence of an island chain}
\label{sec:results}
\subsection{Outside the island chain}
\label{sec:outside}
We now move our attention to a tokamak plasma with $a=1$, $R_0=3$ and $B_0=1$. 
For demonstration purpose, we choose to study an island with parameters $q'=4, \psi_0=(0.5)^2/2=0.125, m_0=5, n_0=2, q_0=5/2$. 
We choose to set $A=10^{-5}$ and $10^{-4}$, corresponding to an island width being $3.2\%$ and $10\%$ of the minor radius, respectively.
Such island widths were observed experimentally in e.g. KSTAR experiments with a tearing mode~\cite{Kwak2013}.
However, the results will apply equally well to islands with a different set of parameters.

\begin{figure}[htbp]
    \centering
    \begin{tabular}{c c}
    \includegraphics[width=7.5cm]{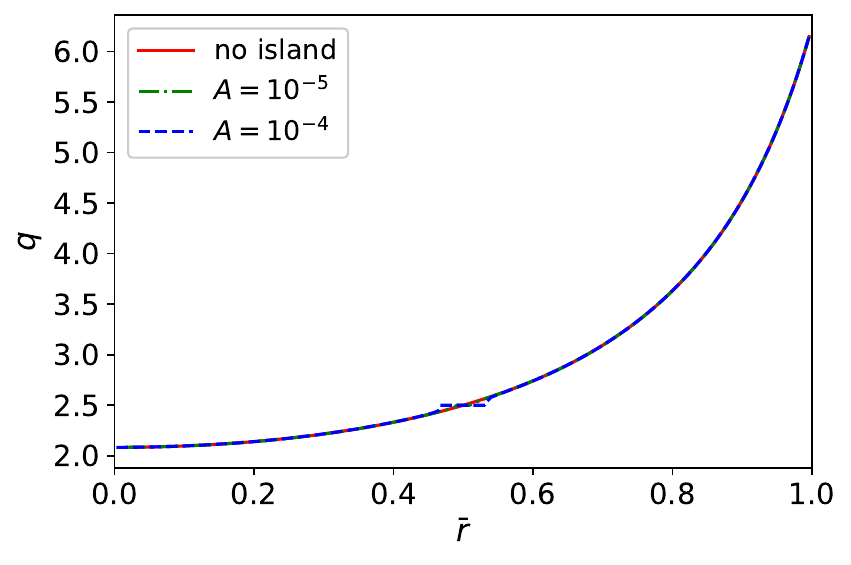}     &
    \includegraphics[width=7.5cm]{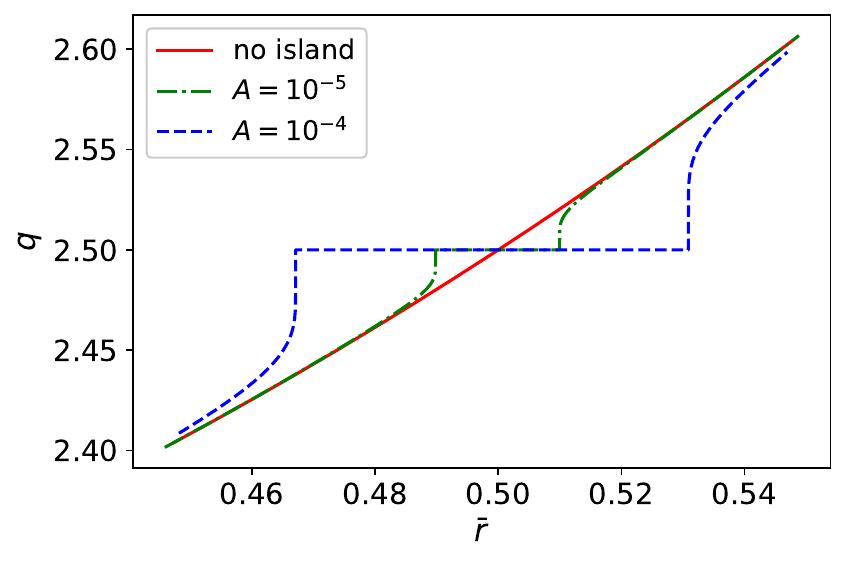} \\
     (a)    &  (b)
    \end{tabular}
    \caption{The $q$ profile with respect to the magnetic axis as a function of the flux label $\bar{r} = \sqrt{\bar{\psi}/\bar{\psi}_{edge}}$ for three cases: and $A=0$ (no island), $A=10^{-5}$ and $A=10^{-4}$. 
    The $q$ profile is plotted for (a) the entire minor radius and (b) the region around the island chain. }
    \label{fig:q_profile}
\end{figure}

We have plotted the $q$ profile with respect to the magnetic axis as a function of the outside flux label $\bar{r} = \sqrt{\bar{\psi}/\bar{\psi}_{edge}}$ for the three cases: and $A=0$ (no island), $A=10^{-5}$ and $A=10^{-4}$,
as shown in \figref{fig:q_profile} (a).
The general shape of the $q$ profiles are determined by our choice of the rotational transform being a linear function of $\psi$, the resonance location $\psi_0$ and the shear $q'$.
The three $q$ profiles are almost identical across the entire minor radius except in the vicinity of the island chain at $\bar{r} = 0.5$,
where a local flattening happens due to the existence of the island in cases where $A\ne 0$.
\Figref{fig:q_profile} (b) zooms into the vicinity of the island chain.
It is evident that the $q$ profile becomes very steep close to the island and then is kept at a constant $q_0$ across the island.

\begin{figure}[htbp]
    \centering
    \begin{tabular}{c c}
    \includegraphics[width=7.5cm]{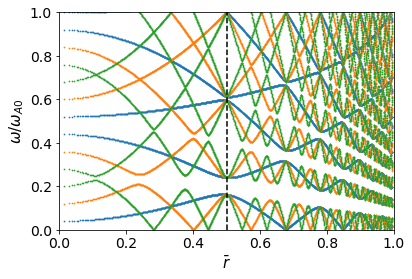} & 
    \includegraphics[width=7.5cm]{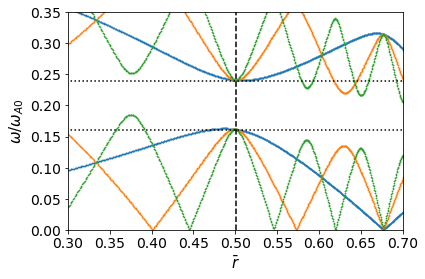}\\
    (a) & (b)   
    \end{tabular}
    \caption{(a) The continuum of the configuration with no island with $n=\pm 1$ (blue), $\pm 3$ (orange), and $\pm 5$ (green). 
    The vertical dash line indicates the location of the island chain in later calculations.
    (b) The same plot zoomed into the range of the TAE frequency and the $q=q_0=5/2$ surface. 
    The two dotted horizontal lines label the tip frequencies of the TAE gap at $\bar{r}=0.5$.}
    \label{fig:continuum_noisland}
\end{figure}

The field period of the island chain is $n_0$, 
and therefore in the case $n_0=2$ all the odd ($n=1$ family) toroidal modes are coupled,
with no coupling to the even modes ($n=0$ family).
We choose to study the $n=1$ family since they form the TAE gap exactly at the location of the island.
The continuum of the configuration with no island is shown in \figref{fig:continuum_noisland} (a) with $n=\pm 1, \pm 3$ and $\pm 5$ and $m=-20$ to $20$.
In the absence of islands, $n$ is a good quantum number.
It is therefore possible to separate the continuum for different $n$'s: they are labelled by different colors in the figure.
One can identify several frequency gaps with increasing frequencies: 
the TAE gap,
the EAE gap,
the non-circularity (triangularity) induced \Alfven eigenmode (NAE) gap, etc.
In this paper, we will focus on the TAE gap, which is formed by the coupling of the branches with poloidal mode number $m$ and $m+1$ due to the toroidal variation of the field strength and metrics,
when the two continuum branches share the same frequency in the cylindrical limit.
\Figref{fig:continuum_noisland} zooms into the range of the TAE frequency and the $q=q_0=5/2$ surface (indicated by the vertical line).
Inspection of the figure shows that the central frequency and the full width of the gap are independent of the toroidal mode number $n$, and are consistent with the theoretical predictions  on the $q=q_0=5/2$ surface~\cite{Cheng1986} given by
\begin{equation}
    \omega_{\text{TAE}} = \frac{1}{2q_0}\omega_{A0},
    \label{eq:w_TAE_outside}
\end{equation}
and
\begin{equation}
    \Delta \omega_{\text{TAE}} = 2 \left(\epsilon_0 + \Delta' \right) \omega_{\text{TAE}},
    \label{eq:width_TAE_outside}
\end{equation}
respectively, where $r_0 = \sqrt{2 \psi_0/B_0}$ is the radius of the island centre and $\epsilon_0 = r_0/R_0$.
Substituting our parameters into \eqref{eq:w_TAE_outside} and \eqref{eq:width_TAE_outside}
yields $\omega_{\text{TAE}} = 0.2 \omega_{A0}$ and $\Delta \omega_{\text{TAE}} = 0.083 \omega_{A0}$.

\begin{figure}[htbp]
    \centering
    \begin{tabular}{c c}
    \includegraphics[width=7.5cm]{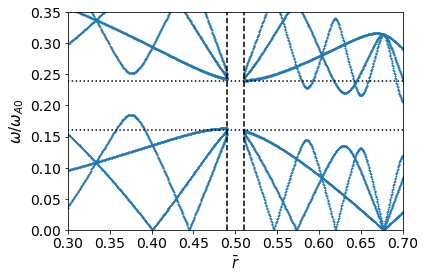} & 
    \includegraphics[width=7.5cm]{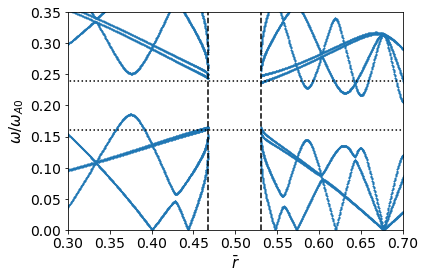}\\
    (a) & (b)   
    \end{tabular}
    \caption{The continuum outside the island chain for (a) $A=10^{-5}$ and (b) $A=10^{-4}$. 
    The vertical dash line indicates the island separatrix. 
    The two dotted horizontal lines label the TAE gap frequencies calculated in the no-island limit,
    same as \figref{fig:continuum_noisland} (b).}
    \label{fig:continuum_island_out}
\end{figure}

The continuum for $A=10^{-5}$ and $A=10^{-4}$
are shown in \figref{fig:continuum_island_out} (a) and (b),
respectively.
With an island chain
$n$ is no longer a good quantum number and therefore we
do not distinguish continuum for different $n$'s.
In the case with a smaller island ($A=10^{-5}$),
there is no notable difference between \figref{fig:continuum_noisland} (b) and \figref{fig:continuum_island_out} (a),
except that the island region is left blank in the latter.
The effect of a larger island ($A=10^{-4}$) is more visible: new gaps are created at around $\bar{r}=0.44$ and $0.56$ 
as a consequence of the broken toroidal symmetry in the vicinity of the island,
namely the HAE gap.
The continuum frequencies of $n=1$ and $n=-1$ branches are also slightly different in \figref{fig:continuum_island_out} (b),
while these frequencies are identical without the island chain.
In both cases, the central frequency and the width of the TAE gap are left mostly unmodified.

\subsection{Inside the island chain}
\label{sec:inside}
\modi
Unlike the flux surfaces outside the island that are only slightly perturbed,
the topological structure of those inside the island chain is completely changed.
They are nested around the island's axis (O point), which should be viewed as the new magnetic axis of the island-stellarator.
\norm
The poloidal asymmetry of the island comes from its strong elongation,
while its toroidal asymmetry comes from the helical winding of the island axis around the core of the tokamak.
Moving along the island axis in the toroidal direction, it alternates between the high-field side and low-field side, resulting in a toroidal modulation of $B$.
Moreover, the flux surfaces are more compressed on the high-field side than on the low-field side due to the Shafranov shift, affecting the island width in real space and hence the metrics.
Besides, the long axis of the island rotates as the island winds around the plasma core,
leading to the helical modulations of $B$ and the metrics.

\begin{figure}[htbp]
    \centering
    \includegraphics[width=8cm]{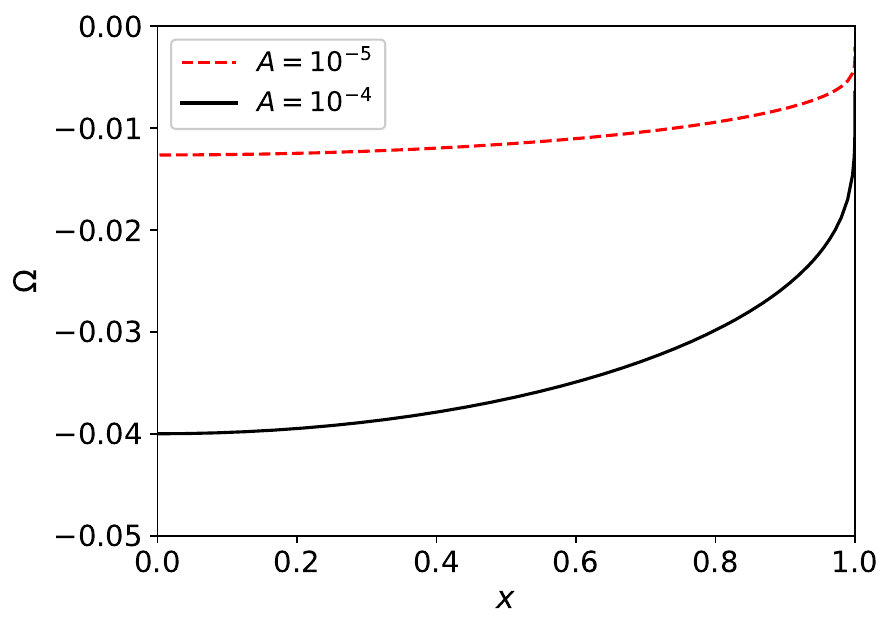}
    \caption{The rotational transform with respect to the island axis as a function of flux label $x=\sqrt{\barpsi/\barpsi_{\text{isl}}}$ inside the island chain.}
    \label{fig:Omega}
\end{figure}

\begin{figure}[htbp]
    \centering
    \begin{tabular}{c c}
    \includegraphics[width=7.5cm]{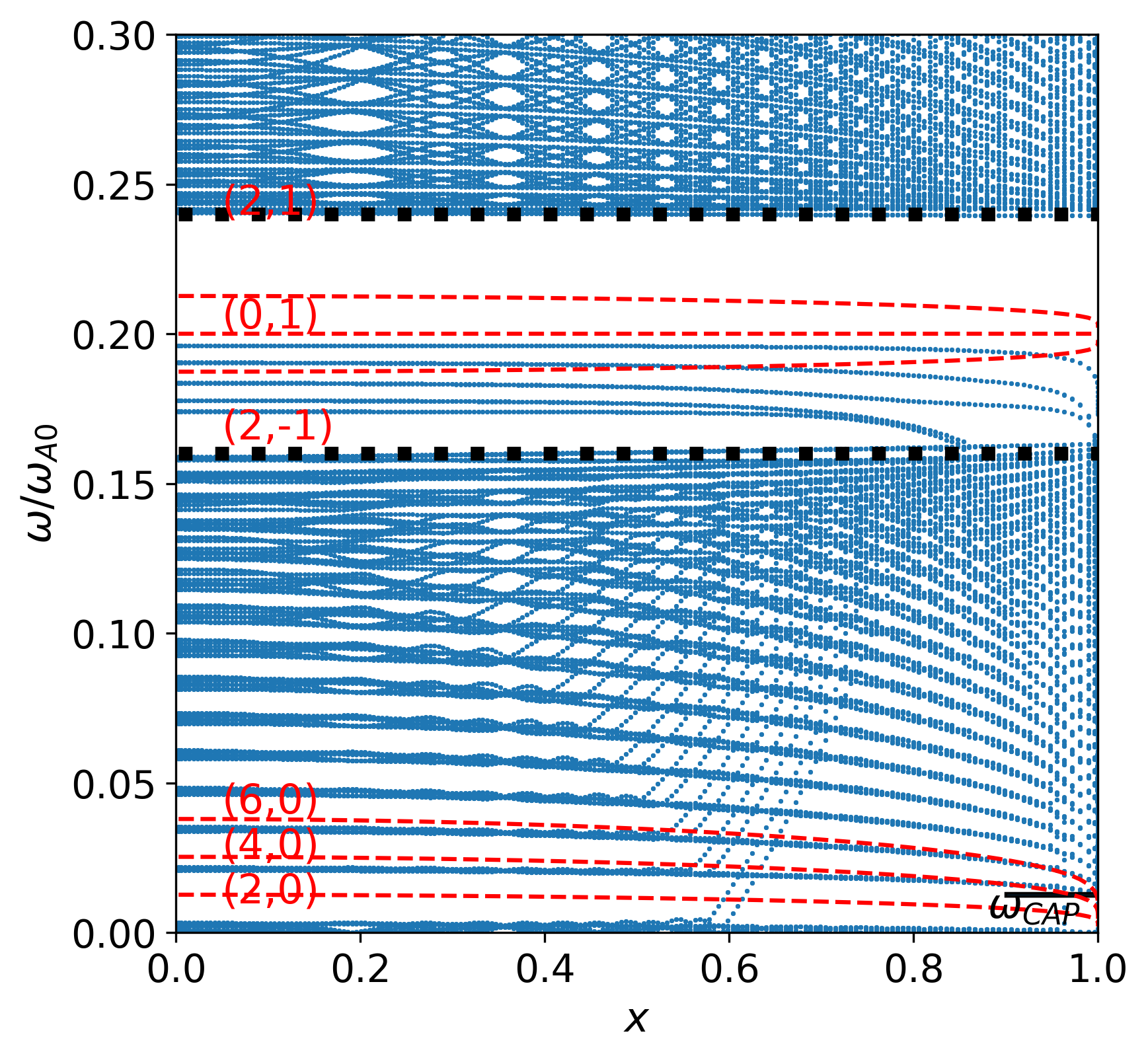} & 
    \includegraphics[width=7.5cm]{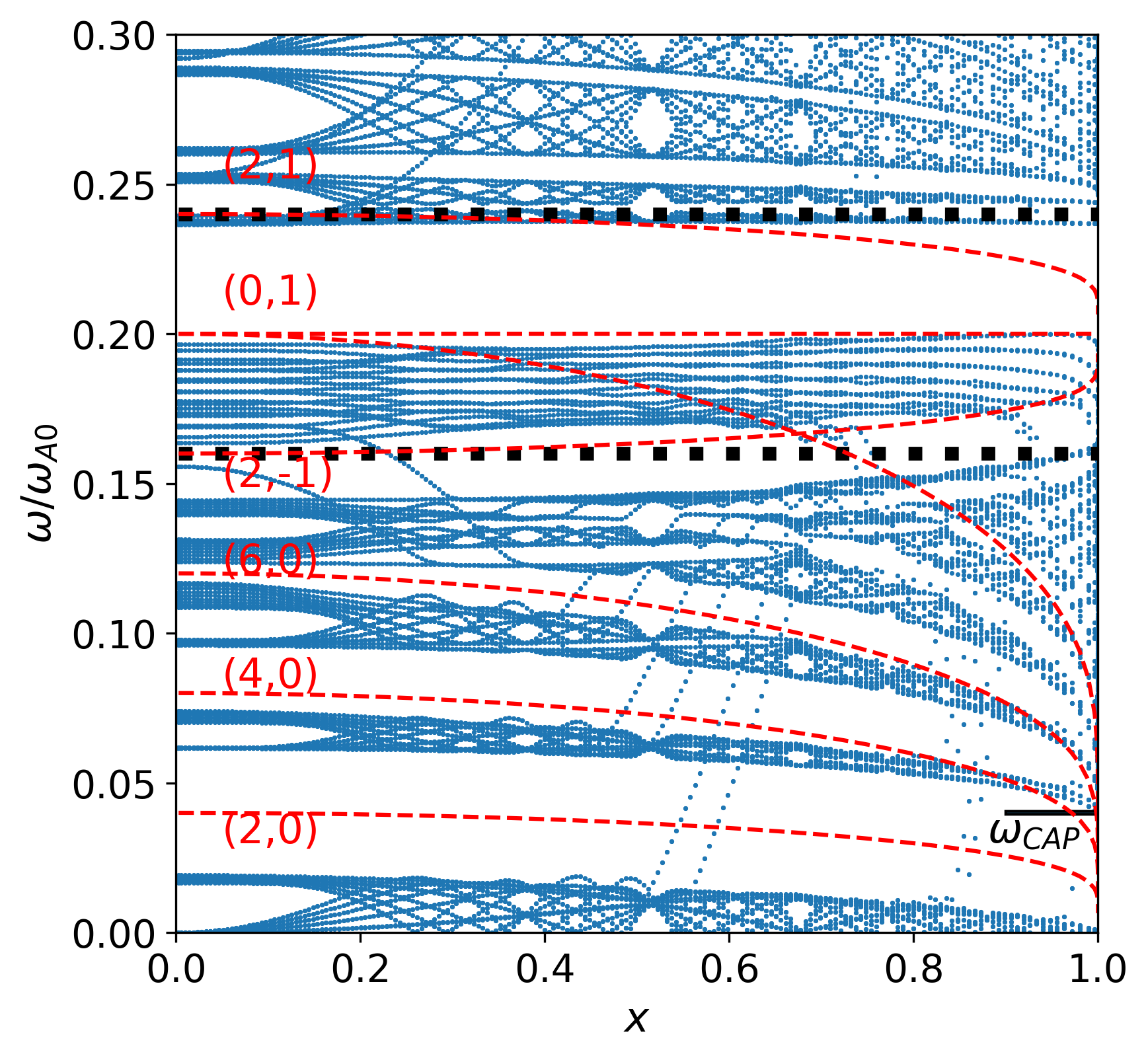}\\
    (a) & (b)   
    \end{tabular}
    \caption{The continuum of the $n=1$ family inside the island chain for (a) $A=10^{-5}$ and (b) $A=10^{-4}$,
    overplotted with gap centre frequencies in \eqref{eq:gap_central_fqc}. 
    The two dotted horizontal lines label the outside TAE gap frequencies in the no-island limit,
    same as \figref{fig:continuum_noisland} (b).
    The solid black line labels $\omega_{\text{CAP}}$ computed in \eqref{eq:w_CAP}.
    The range of Fourier harmonics are $m=90$ to $90$ and $n=-7$ to $7$ with an increment of $2$ for (a), and $m=-40$ to $40$ and $n=-11$ to $11$ with an increment of $2$ for (b)}
    \label{fig:continuum_island_in}
\end{figure}

We have shown in \figref{fig:Omega} the rotational transform $\Omega$ with respect to the island axis as a function of $x=\sqrt{\barpsi/\barpsi_{\text{isl}}}$ for $A=10^{-5}$ and $A=10^{-4}$,
while all other parameters are the same as \secref{sec:outside}.
Similar to the outside, the continuum inside the island chain also has a $n=1$ family and a $n=0$ family.
\modi
We note that the flat $q$ profile across the island in \figref{fig:q_profile} is computed with respect to the original magnetic axis and is defined as the average over an infinite number of turns. The differences between flux surfaces inside an island are averaged out. On the other hand, the rotational transform in \figref{fig:Omega} is computed with respect to the island axis. 
As an analogy, the former describes how a power cable twists on the surfaces of a toroid as a whole (so it’s the same for each individual wire), while the latter describes how each wire inside twists around the cable centre.
\norm
We again choose to study the $n=1$ family as in the cylindrical limit,
the frequencies of the $n=1$ modes approach the centre of the TAE gap outside.
We have computed the continuum frequencies for the $n=1$ family inside the island chain for both choices of $A$ shown in \figref{fig:continuum_island_in} (a) and (b), respectively.
With all the aforementioned modulations at similarly strong amplitudes, the continuum has an extensive zoo of frequency gaps.
It is also noteworthy that close to the separatrix the rotational transform is very close to zero, requiring a much higher, if not infinite, number of Fourier harmonics~\cite{Konies2022}.
In our calculation, we have chosen the number of Fourier harmonics to resolve the continuum near the O point more accurately, which leads to some numerical artifacts at, for example, around $x=0.6$ inside the gaps in both figures due to an insufficient numerical resolution. However, the overall picture of the continuum is unchanged.

We assume a small gap width and ignore interactions between different gaps, i.e. considering the limit that the full gap width $\Delta \omega^{(\Delta m, \Delta n)}$ is much smaller than the distance from neighbouring gaps and $\omega^{(\Delta m, \Delta n)}$ itself.
In the cylindrical limit, the continuum frequency of branch $(m,n)$ is given by
\begin{equation}
    \frac{\omega(\Omega)}{\omega_{A0}} = \left| m \Omega  - \frac{n}{m_0} \right|.
    \label{eq:island_cylindrical_omega}
\end{equation}
The additional $m_0$ in the denominator is due to the periodicity of the toroidal angle being $2m_0 \pi$, as stated in \secref{sec:theory_continuum}.
When two branches $(m,n)$ and $(m+\Delta m,n + \Delta n_0)$ intercepts,
it would mean $\left| m \Omega  - n/m_0 \right| = \left| (m+\Delta m ) \Omega  - (n + \Delta n n_0)/m_0 \right|$,
leading to $\Omega$ at the interception being
\begin{equation}
    \Omega = \frac{2n + \Delta n n_0}{m_0(2m + \Delta m)},
    \label{eq:Omega_m_n}
\end{equation}
which, when substituted back to \eqref{eq:island_cylindrical_omega} to eliminate $n$, yields
\begin{equation}
    \omega^{(\Delta m, \Delta n)} = \frac{1}{2}\left|\Delta n \frac{n_0}{m_0} - \Delta m \Omega \right| \omega_{A_0},
    \label{eq:gap_central_fqc}
\end{equation}
for the central frequency of the $(\Delta m, \Delta n)$ gap.

Using \eqref{eq:gap_central_fqc} and the $\Omega$ profile in \figref{fig:Omega},
we have computed $\omega^{(\Delta m, \Delta n)}$ for the following gaps:
$(2,0)$, $(4,0)$, $(6,0)$, $(2,-1)$ and $(0,1)$.
These frequencies as a function of $x$ are added to \figref{fig:continuum_island_in}, helping us to identify the gaps.
Since the gaps are clearer and wider in the $A=10^{-4}$ case,
we list them from low to high frequency in \figref{fig:continuum_island_in} (b) as follows.
For most of the gaps the gap centre frequencies in \eqref{eq:gap_central_fqc} match well with the figure.
\begin{itemize}
    \item \modi The island TAE $(1,0)$ gap is negligibly small.
    Its width is linked to the inverse aspect ratio of the island, defined by
    \begin{equation}
        \epsilon_{\text{isl}} = \frac{\Delta r}{R_0},
    \end{equation}
    in which $\Delta r = w / (2 r_0 B_0)$ is the island half width in radius.
    For $A=10^{-4}$, $\epsilon_{\text{isl}}$ is on the order of $1\%$.
    The island TAE gap is therefore significantly smaller than the predictions given by Biancalani \etal~\cite{Biancalani2010prl, Biancalani2010}, where the inverse aspect ratio of the whole machine was used to estimate its width.
    \norm
    \item The EAE $(2,0)$ gap (alias the MiAE gap) is created by the strong ellipticity of the island.
    \modi
    This can be described by the ratio between the island's short axis and long axis and is given by
    \begin{equation}
        e = \frac{\Delta r} { 2 \pi r_0 / (2 m_0)} \approx 0.13,
    \end{equation}
    for $A=10^{-4}$. 
    \norm
    On the island axis $\Omega = -m_0 \Omega_0$,
    so \eqref{eq:gap_central_fqc} gives $\omega^{(2,0)} = |-m_0 \Omega_0|= n_0 \sqrt{A q'}$ with $\Delta m =2$, identical to $\omega_{\text{CAP}} = n_0 \sqrt{A q'}$ in \eqref{eq:w_CAP}.
    On the island separatrix, the upper frequency of the gap approaches $\omega_{\text{CAP}}$ as indicated in \figref{fig:continuum_island_in} (a).
    This would guarantee the EAE/MiAE gap to be an open gap inside the island chain. A further global calculation is needed to find a discrete MiAE within this gap.
    \item The $(4,0)$ gap and the $(6,0)$ gaps are the next two wide gaps, even when the corresponding $m=4,6, n=0$ Fourier harmonics of the metric and field are small.
    In this case, the $(m,n)$ mode does not couple directly with the $(m+4,n)$ mode to create the gap, for example, but through the coupling to the $(m+2,n)$ mode first via the $m=2, n=0$ harmonic and then to the $(m+4,n)$ mode.
    The upper and low frequencies of both gaps converge to $\omega_{\text{CAP}}$ on the separatrix.
    \item At around $\omega/\omega_{A0} = 0.15$ on axis sits the $\text{HAE}_{2,-1}$ $(2,-1)$ gap due to the helical modulations.
    \modi
    Unlike other gaps, the gap frequency goes up with $x$.
    \norm
    This is a new gap in toroidal geometry.
    Between the $(6,0)$ and $(2,-1)$ gap, the $(2M,0)$ gap family and the $(2M,-1)$ gap family clash and cross each other, creating very complicated patterns.
    The clash leads to a reduction of the gap width or even a complete elimination of one or both gaps at the crossing site~\cite{Yakovenko2007}.
    \item The mirror-induced \Alfven eigenmode (MAE) $(0,1)$ gap has a constant frequency at  \begin{equation}
        \omega_{\text{MAE}} = \frac{1}{2q_0} \omega_{A0}.
    \end{equation}
    This frequency is identical to the TAE gap frequency outside the island chain at $q=q_0$
    in \eqref{eq:w_TAE_outside}.
    Therefore the MAE gap can be viewed as the continuation of the outside TAE gap into the island chain. 
    The gap width is however narrower than the outside TAE gap plotted in black. We will take a closer look at the MAE gap in \secref{sec:MAE}.
    \item There are other smaller gaps of the $(2M,1)$ family beyond the MAE gap.
\end{itemize}
Comparing \figref{fig:continuum_island_in} (a) to (b),
one shows that the $(\Delta m, 0)$ gaps are narrower and at lower frequencies, if the island size is smaller.
This is consistent with the prediction of Biancalani \etal as the frequencies of the lower part of the spectrum scales with $\sqrt{A}$ the island width.
The MAE and HAE gaps, however, are not changing dramatically, as they are related to the toroidicity of the machine independent of the size of the island.
The gap centre frequencies in \eqref{eq:gap_central_fqc} are no longer accurate for the HAEs due to the large MAE gap and the small gap width assumption made in deriving it becoming invalid.

In fact, \figref{fig:continuum_island_in} shares a great similarity to continuum in Helias configurations~\cite{Kolesnichenko2001} (e.g. stellarators of Wendelstein line),
where a combination of EAE gap, MAE gap and HAE gaps is present.
The main difference is that the $\text{HAE}_{21}$ and $\text{HAE}_{22}$ gaps in Helias are much wider than other gaps, while the gaps with a negative $\Delta n$, e.g. $\text{HAE}_{2,-1}$, are absent.
In an island chain, the widest gaps are the MAE gap and the $\text{HAE}_{2,-1}$ gap, or the EAE gap when the island size is more significant.

\subsection{The MAE gap}
\label{sec:MAE}
\begin{figure}[htbp]
    \centering
    \includegraphics[width=8cm]{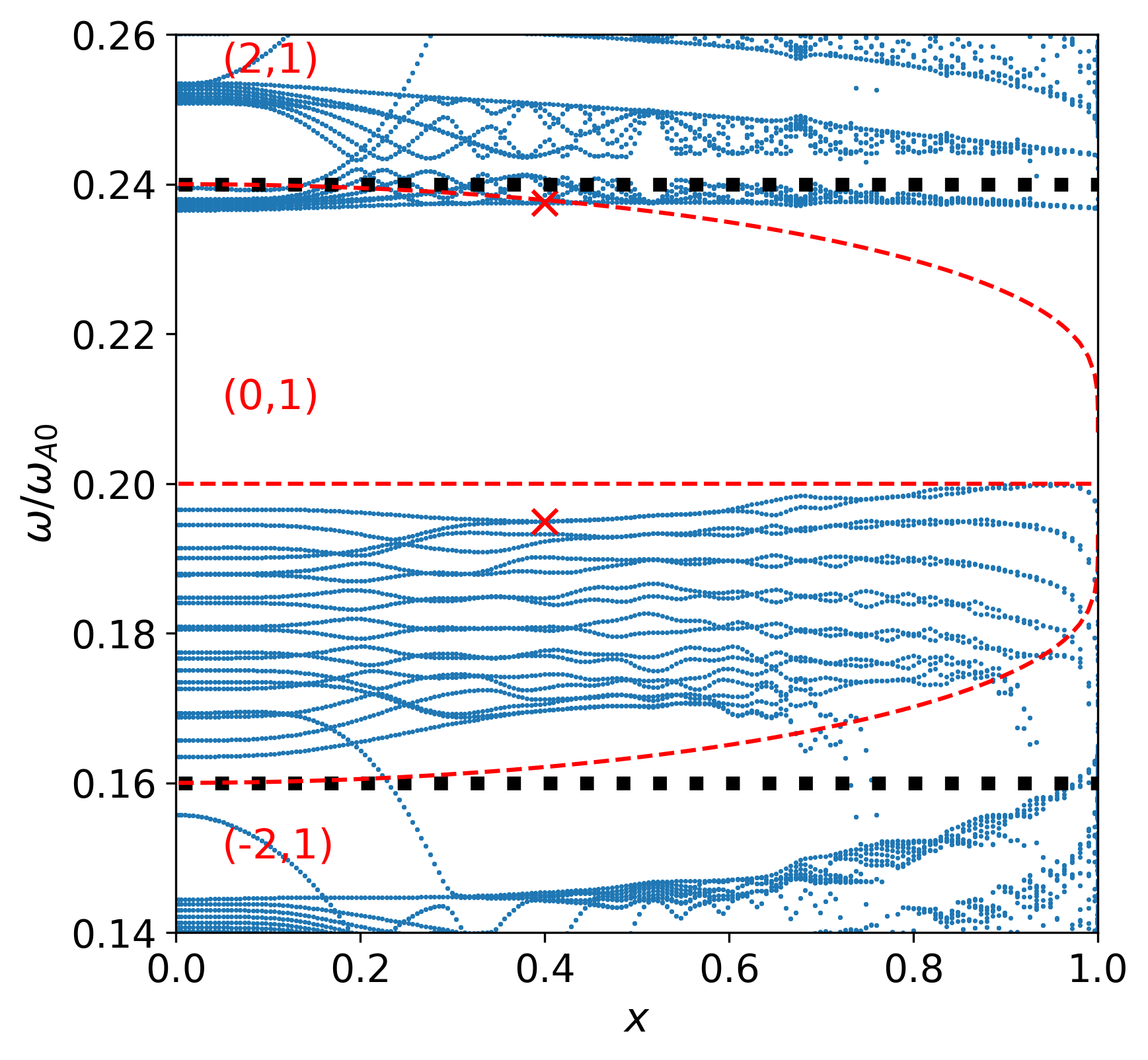}
    \caption{The continuum of the $n=1$ family inside the island chain for $A=10^{-4}$ zoomed into the MAE frequency region,
    overplotted with gap centre frequencies in \eqref{eq:gap_central_fqc}. 
    The two dotted horizontal lines label the outside TAE gap frequencies in the no-island limit.
    The two ``x'' markers at $x=0.4$ indicate the two eigenfunctions being studied in \secref{sec:localised}. }
    \label{fig:continuum_island_zoom}
\end{figure}

\begin{figure}[htbp]
    \centering
    \includegraphics[width=8cm]{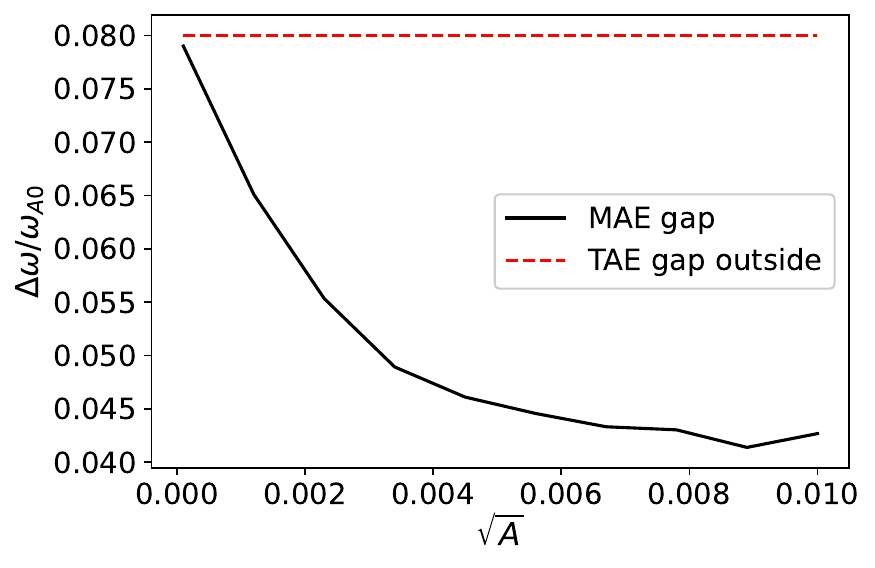}
    \caption{The width of the inside MAE gap measured at $x=0.4$ inside the island as a function of $\sqrt{A}$ which is proportional to the island width.
    A comparison with the TAE gap width outside the island is also given.}
    \label{fig:MAE_width}
\end{figure}

As mentioned in \secref{sec:inside}, the MAE gap can be viewed as the continuation of the outside TAE gap into the island chain.
It can therefore impact a global TAE and will be our main focus.
\Figref{fig:continuum_island_zoom} zooms into the vicinity of the MAE gap for the $A=10^{-4}$ case, overplotted with the outside TAE gap in thick black lines. 
Clearly, \figref{fig:continuum_island_zoom} demonstrates an up-shift of the lower tip compared to the TAE gap outside,
leading to an overall shrinkage of the MAE gap width.
We have measured the gap width at $x=0.4$ and plotted it in \figref{fig:MAE_width} as a function of $\sqrt{A}$, which is proportional to the island width.
The gap width shrinks as the island width increases and then saturates at around half of the original width of the TAE gap.

Again assuming small and non-interacting gaps,
the full width of the $(\Delta m, \Delta n)$ gap can be computed analytically by retaining only Fourier modes with mode numbers $(m,n)$ and $(m+\Delta m,n + \Delta n_0)$
and solving \eqref{eq:continuum} at the interception,
giving that
\begin{equation}
    \Delta \omega^{(\Delta m, \Delta n)} \approx \left| 2 \epsilon_B^{(\Delta m, \Delta n)} -\epsilon_g^{(\Delta m, \Delta n)}\right|\omega^{(\Delta m, \Delta n)},
    \label{eq:gap_width}
\end{equation}
where $\epsilon_B^{(m, n)}$ and $\epsilon_g^{(m, n)}$ are the Fourier coefficients of $B$ and $|\gradbarpsi|^2$, respectively so that
\begin{equation}
    B = B_0 \left[ 1 + \sum_{m,n} \epsilon_B^{(m, n)} \cos \left(m \baralpha - \frac{n n_0}{m_0} \zeta \right) \right],
\end{equation}
and
\begin{equation}
    |\gradbarpsi|^2 = |\gradbarpsi|^2_0 \left[ 1 + \sum_{m,n} \epsilon_g^{(m, n)} \cos \left(m \baralpha - \frac{n n_0}{m_0} \zeta \right) \right].
\end{equation}
A full derivation of \eqref{eq:gap_width} can be found in Kolesnichenko \etal~\cite{Kolesnichenko2001} and will not be repeated here. Next, we need to compute $\epsilon_B^{(m, n)}$ and $\epsilon_g^{(m, n)}$.

Around the island axis where $\kappa \ll 1$, the Hamiltonian \eqref{eq:chi_quadratic} resembles that of a harmonic oscillator, with the relationship between $\vartheta$, $\psi$ and $\baralpha$
simply given by
\begin{equation}
\vartheta = \alpha + \frac{\zeta}{q_0} \approx  \frac{2 \sqrt{\kappa}}{m_0} \sin \baralpha + \frac{\zeta}{q_0},
\label{eq:theta_reduced}
\end{equation}
and
\begin{equation}
    \psi \approx \psi_0 + \frac{w}{2} \sqrt{\kappa} \cos \baralpha.
    \label{eq:psi_reduced}
\end{equation}
To the order $O(\epsilon)$, one has $B \approx B_0 (1 - r/R_0 \cos \vartheta)$ and therefore,
after substituting in \eqref{eq:theta_reduced} and \eqref{eq:psi_reduced},
one reaches the limit $\kappa \rightarrow 0$ that
\begin{equation}
    B \approx B_0 \left(1 - \epsilon_0 \cos \frac{n_0 \zeta}{m_0} \right).
\end{equation}
This gives $\epsilon_B^{(0,1)} = - \epsilon_0$ and $\epsilon_B^{(m, n)} = 0$ for other combinations of $(m,n)$. Similarly, one can substitute \eqref{eq:theta_reduced} and \eqref{eq:psi_reduced}
into \eqref{eq:gradpsi2} and take the same limit $\kappa \rightarrow 0$.
After some algebra (see \ref{app:gradpsi} for details), one obtains that 
\begin{equation}
    \epsilon_g^{(0,1)} = \frac{2}{1 + \delta^2} \left[ \Delta' - \delta^2\left( \epsilon_0 + \Delta' \right) \right],
\end{equation}
with $\Delta'$ evaluated at $r = r_0$ and $\delta = A m_0 / (\Omega_0 B_0 r_0^2) $.
The estimation of the MAE gap width from \eqref{eq:gap_width} is therefore given by
\begin{equation}
    \Delta \omega_{\text{MAE}} \approx 2 \left[  \epsilon_0 +  \Delta' - \frac{\delta^2}{1 + \delta^2} \left( \epsilon_0 + 2 \Delta' \right) \right] \omega_{\text{MAE}}.
    \label{eq:width_MAE}
\end{equation}
The contribution of the island geometry to the gap width is contained in the third term of \eqref{eq:width_MAE}, which acts to reduce it.
In the limit $\delta \rightarrow 0$ (vanishing island size), 
one gets $\Delta \omega_{\text{MAE}} = \Delta \omega_{\text{TAE}}$,
where $\Delta\omega_{\text{TAE}}$ is the TAE gap width outside the island in \eqref{eq:width_TAE_outside}.

However, $\delta$ is generally speaking a small quantity.
\modi
\Eqref{eq:width_MAE} alone cannot explain the much smaller MAE  gap width compared to the TAE gap outside.
\norm
For instance, the estimated gap widths are $0.0827 \omega_{A0}$ and $0.0775\omega_{A0}$ for $A=10^{-5}$
and $A=10^{-4}$, respectively,
while these numbers are $0.0443 \omega_{A0}$ and $0.0400 \omega_{A0}$ by measuring \figref{fig:continuum_island_in} on the island axis.
It would mean that the assumption used to derive \eqref{eq:width_MAE}, i.e. the coupling terms are small and the gaps are far away without interactions,
is no longer valid.

We conjecture that the smaller MAE gap is a consequence of the competition between the MAE gap and the $\text{HAE}_{2,-1}$ gap.
Around the island axis, one obtains that
\begin{equation}
    \epsilon_g^{(2,-1)} = \frac{1}{1+ \delta^2} \left\{\Delta' + \delta [\epsilon_0 + (r\Delta')'] + \delta^2(\epsilon_0 + \Delta') \right\},
\end{equation}
which is on the same order as $\epsilon_g^{(0,1)}$ and $\epsilon_B^{(0,1)}$, meaning that the impact of the $\text{HAE}_{2,-1}$ gap on the MAE gap is not negligible when the two gaps are very close. 
A closer look at the Fourier coefficients $\epsilon_g^{(m,n)}$ also reveals that on the island axis,
\begin{equation}
    \epsilon_g^{(2,0)} = \frac{1 - \delta^2}{1 + \delta^2},
\end{equation}
which is on the order of unity.
For $A = 10^{-4}$, one has $\epsilon_g^{(2,0)}=0.88$.
The Fourier harmonic $(m,n)$ is therefore strongly coupled to
all $(m + 2 M,n)$ for any integer $M$ through $\epsilon_g^{(2,0)}$, and then to the class $(m + 2 M,n \pm N)$ for a wide range of integers $N$.
It would mean that one can no longer treat the MAE gap as the consequence of coupling between only a few Fourier harmonics with similar frequencies but should take into account all Fourier harmonics in the calculation.

\begin{figure}[htbp]
    \centering
    \includegraphics[width=8cm]{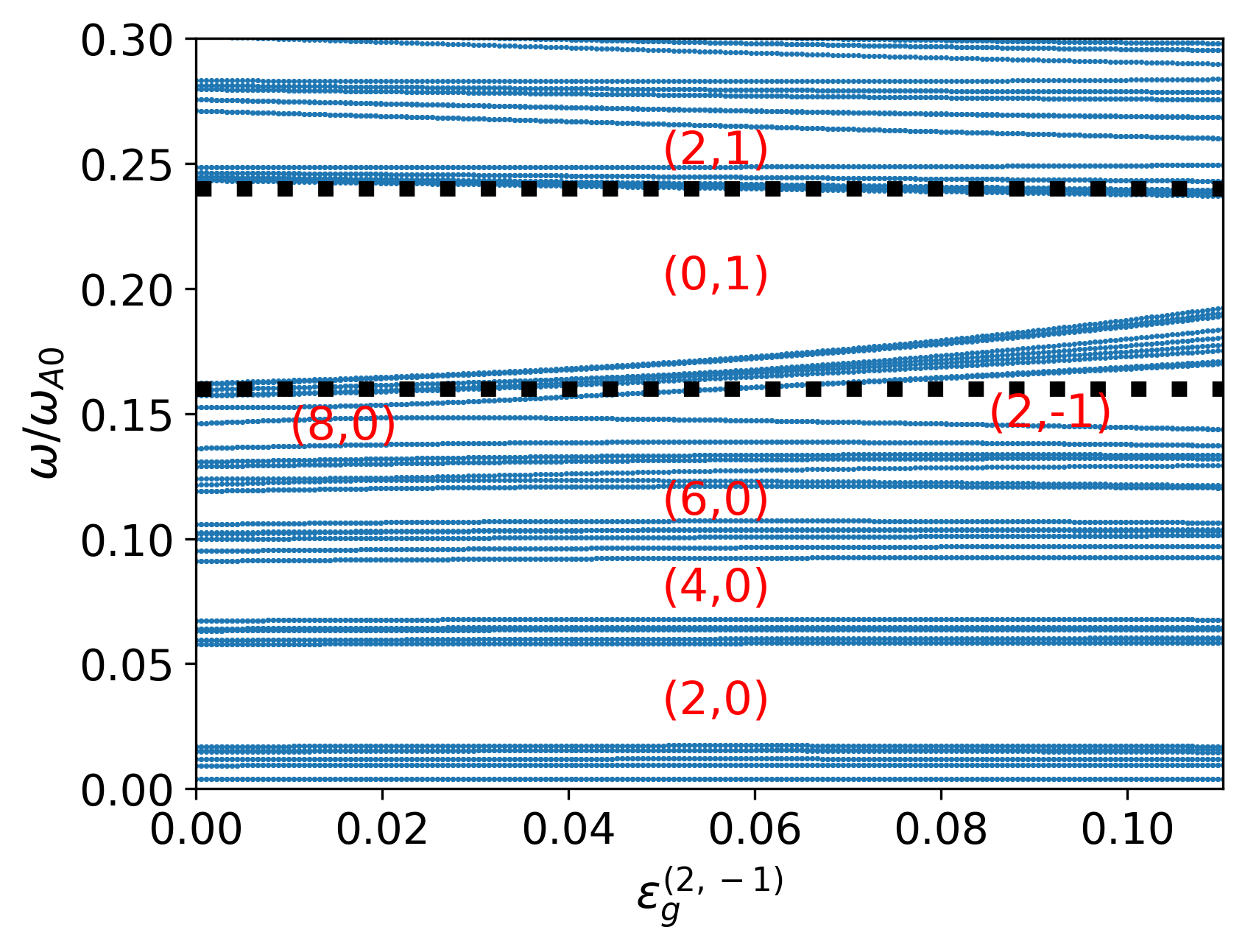}
    \caption{The continuum spectrum of the $n=1$ family inside the island chain at $x=0.4$ for $A=10^{-4}$ as a function of $\epsilon_g^{(2,-1)}$. 
    The locations of the gaps are indicated by their $(\Delta m, \Delta n)$.
    The two dotted horizontal lines label the outside TAE gap frequencies in the no-island limit.}
    \label{fig:continuum_scan_eps}
\end{figure}

To verify our conjecture, we compute numerically the continuum spectrum at $x=0.4$ for $A=10^{-4}$.
In our scan, we scale back the value of $\epsilon_g^{2,-1}$ starting from zero while keeping $B$ and all other harmonics of $|\gradbarpsi|^2$ unchanged, as shown in \figref{fig:continuum_scan_eps}.
When $\epsilon_g^{(2,\pm 1)} = 0$, the $(2,-1)$ gap is closed and its space is occupied by the $(8,0)$ gap. 
At this point, one has an agreement between the MAE gap width and that of the outside TAE gap.
When $\epsilon_g^{(2,\pm 1)}$ gradually increases,
the $(2,-1)$ gap opens and pushes its wall with the MAE gap upwards,
and as such, leads to the shrinkage of the MAE gap width.
A similar behaviour was discovered by Kolesnichenko \etal~\cite{Kolesnichenko2001} in Helias configurations, showing an overall up-shift of the MAE gap in order to avoid overlapping with the HAE gap.

\subsection{Localised continuum modes}
\label{sec:localised}
The impact of toroidal asymmetry is more than just the introduction of new gaps and the complication of their interactions.
The property of the spectra can be fundamentally changed.
That is, \eqref{eq:continuum} can possess not only continuous but also dense discrete spectra~\cite{Salat1992}.
Inspection of \figref{fig:continuum_island_zoom} reveals the existence of a group of micro-gaps between the MAE gap and the $\text{HAE}_{2, -1}$ gap.
The ``walls'' between these gaps as well as the MAE gap are extremely thin.
This is a feature of the continuum with a very high level of asymmetry in both angles:
one can always find a pair of $(\Delta m, \Delta n)$ such that \eqref{eq:gap_central_fqc}
is arbitrarily close to any given frequency.
That is to say, the spectrum approaches a Cantor set~\cite{Salat1992} such that there are opened gaps almost everywhere with non-negligible width and the walls between them become infinitesimally thin.
The spectra, i.e. the gap walls, become dense and discrete,
dubbed as the non-symmetry induced \Alfven eigenmode (NSAE) by Salat and Tataronis~\cite{Salat1992}.
They are similar to the \Alfven gap modes existing in the continuum gaps of the ballooning spectrum, such as the TAEs or EAEs.
However, the TAEs and EAEs are isolated discrete points in the spectrum,
while the NSAEs are dense and discrete with potentially a infinitesimal gap between the neighbours due to the mathematical property of a Cantor set.

\begin{figure}[htbp]
    \centering
    \begin{tabular}{c c}
    \includegraphics[width=7.5cm]{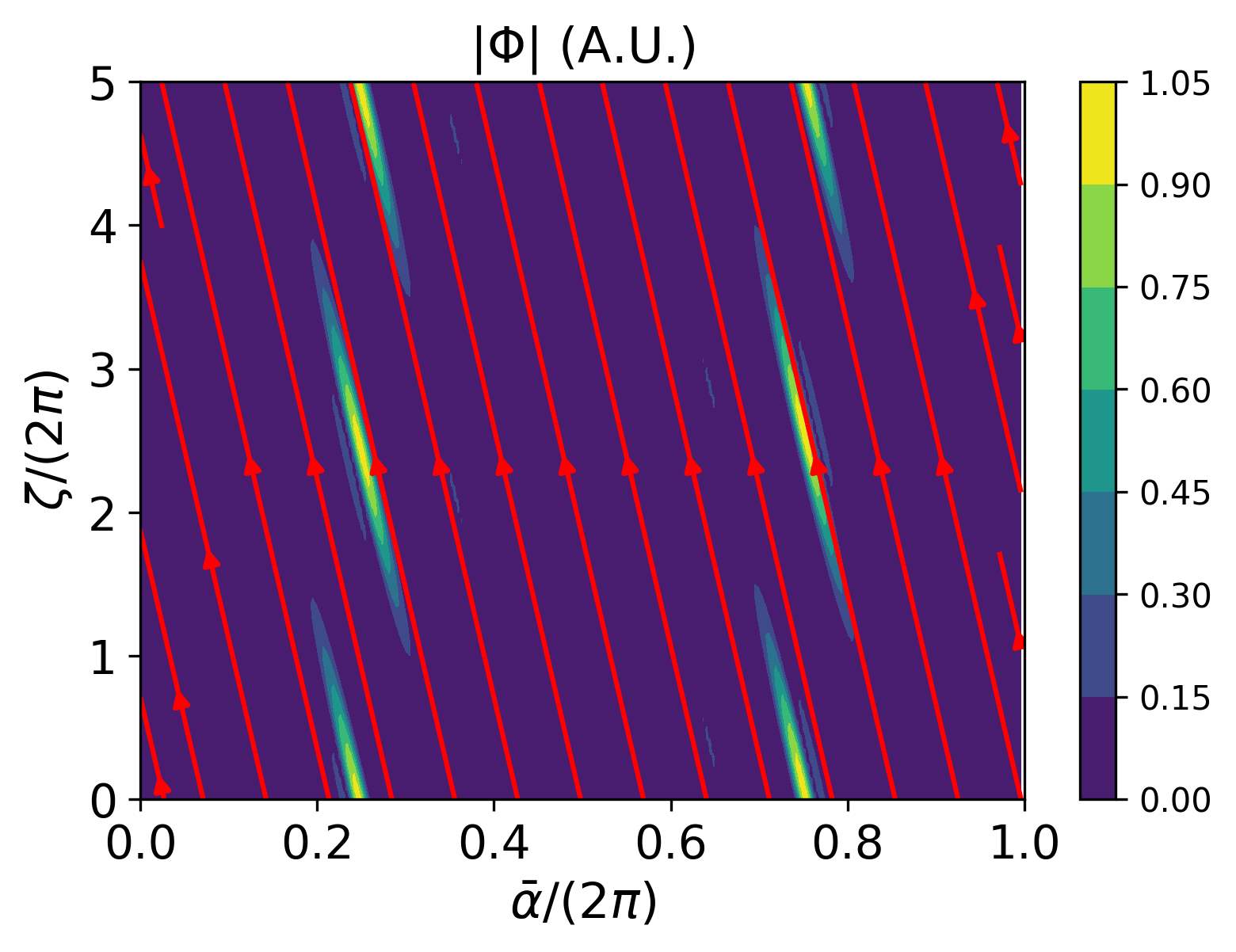} & 
    \includegraphics[width=7.5cm]{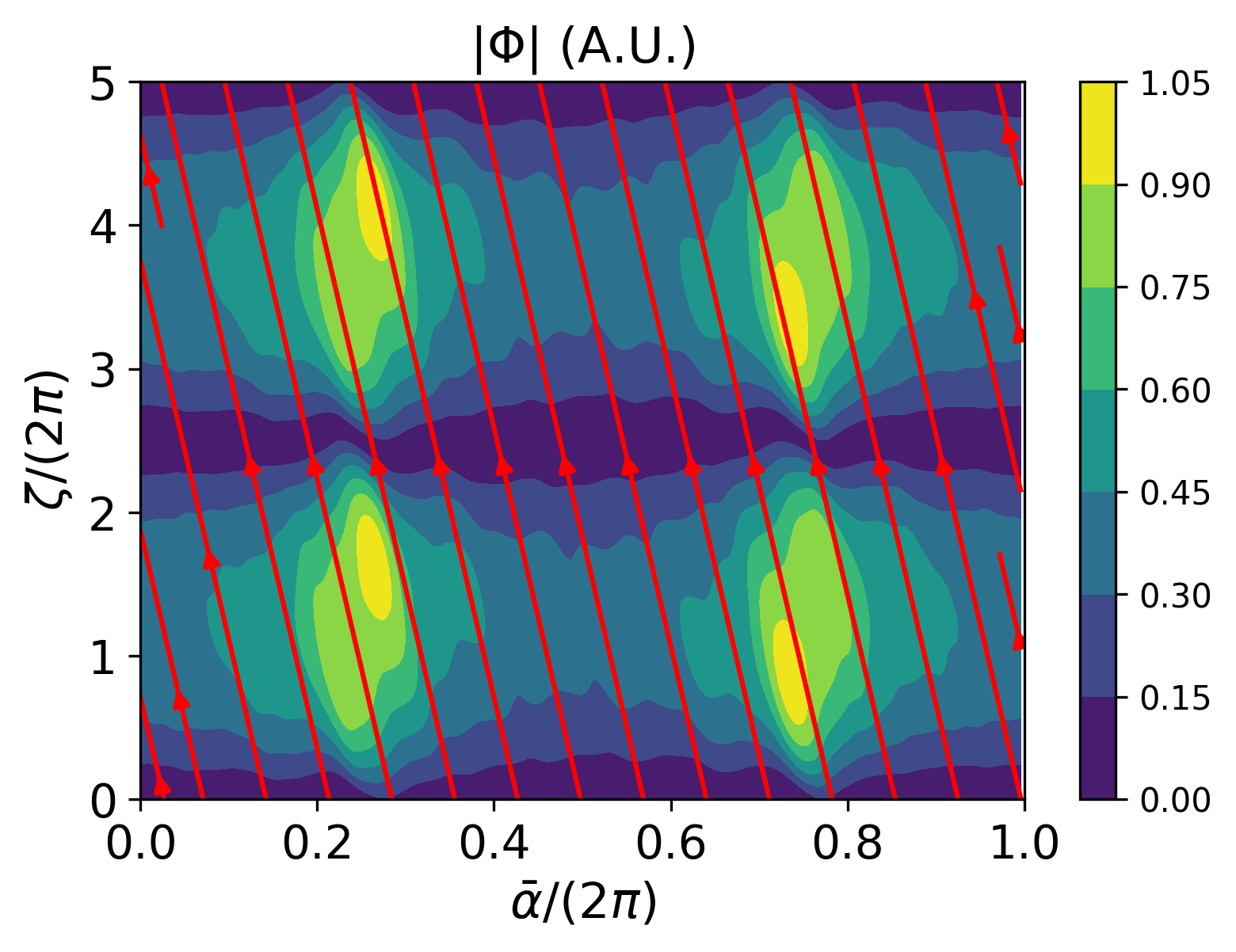}\\
    (a) & (b)   
    \end{tabular}
    \caption{The 2D eigenfunctions of (a) an Anderson localised solution with frequency on the lower tip of the inside MAE gap in \figref{fig:continuum_island_zoom} and (b) a global Bloch solution with frequency on the upper tip, overplotted with the magnetic field lines.}
    \label{fig:eigenfunction}
\end{figure}

Unlike the continuum modes which have global (2D) oscillatory eigenfunctions on the $\baralpha-\zeta$ plane, the eigenfunctions of the NSAEs are highly localised~\cite{Salat1992,Salat1997,Salat2001,Salat2001a}.
In \figref{fig:eigenfunction} (a), 
we have plotted the 2D eigenfunction of a NSAE mode at $x=0.4$ with its frequency on the lower tip of the MAE gap as indicated by the lower ``x'' symbol in \figref{fig:continuum_island_zoom}.
The eigenfunction is localised around $\baralpha=\pi/2$ and $3 \pi/2$,
corresponding to the vicinity of the island long axis.
The mode structure only spans a limited distance along the magnetic field lines around its peak,
while across the field line its amplitude decays rapidly.
Due to their similarities to ballooning modes whose localisation are within the bad curvature region,
NSAEs are also given the name \Alfven-ballooning modes by Salat \etal.
In comparison, we show in \figref{fig:eigenfunction} (b) an oscillatory continuum mode,
whose frequency is on the upper tip of the MAE gap.
The mode structure in this case has a broad span over the entire 2D plane.

To explain the formation of the localised mode structure, 
one can make the coordinate transformation $a = \baralpha - \Omega \zeta$ to replace $\baralpha$.
After such a transformation, the operator $\bdotgrad$ becomes $\bdotgrad = J^{-1} \partial_\zeta$ 
and, with only the $\zeta$ derivative, \eqref{eq:continuum} simplifies to an ordinary differential equation of the variable $\zeta$ on each field line labelled by $a$.
The consequence of this simplification is the loss of the double-periodic boundary condition as well as the double-periodicity of the coefficients on surfaces with an irrational rotational transform $\Omega$, since the field lines no longer close back on themselves.
On a single field line with a fixed $a$, one can write for example
$B(\baralpha, \zeta) = B(a + \Omega \zeta, \zeta)$, and as such, $B$ no longer has any periodicity in $\zeta$ if it has both poloidal and toroidal dependencies.
Instead, $B$ is called quasi-periodic created by the beating of Fourier components with periods of irrational ratios, showing a mixture of regular patterns and randomness.
The continuum equation \eqref{eq:continuum} is now equivalent to the time-independent Schr\"{o}dinger's equation with a quasi-periodic potential well and the eigenvalue problem is now the same as finding an energy level of the corresponding quantum particle system.
In fact, Salat and Tataronis showed that with a proper change of coordinate and eigenfunction,
\eqref{eq:continuum} can indeed be converted to the Schr\"{o}dinger's equation.

In the well-known Schr\"{o}dinger's equation with a periodic potential,
all eigenfunctions take the Bloch wave solution $\Phi(\zeta) = e^{ik\zeta} \phi(\zeta)$
as a consequence of the Bloch theorem in condensed matter physics,
in which $\phi(\zeta)$ has the same periodicity as the system.
It gives a globally propagating wave $e^{ik\zeta}$ modulated by a local shape, corresponding to the solution to \eqref{eq:continuum} should the system possess a toroidal symmetry. 
Dinaburg and Sinai ~\cite{Dinaburg1976} proved mathematically that with a small deviation from periodicity or with a high energy level, the continuous spectra still exist and the solutions remain a Bloch wave.
Our oscillatory continuum solutions belong to this class.
However, when the deviation from periodicity is high enough, 
localised wave solutions, whose amplitude decays exponentially away from the localisation site, are created by a process known as Anderson localisation~\cite{Sinai1987, Frohlich1990}.
In Anderson localisation, wave packets are scattered back from the random potential when they propagate in either direction and try to escape, whose interference creates a strong peak at the localisation site.
This gives a physical explanation of our localised NSAE solutions.
Note that Anderson localised modes are similarly found in the ballooning spectrum of stellarators by Cuthbert and Dewar~\cite{Cuthbert2000}.

\subsection{The combined spectrum}
Now we are in the position to combine the continuum inside and outside the island chain.
They are given in different radial coordinates and therefore cannot be plotted together directly.
Starting from the magnetic axis in \figref{fig:poincare}, one moves outward on the mid-plane until it reaches the inner side of the island.
To reach the outer side, one will need to go across the flux surfaces within the island from $x=1$ to $x=0$, then again from $x=0$ back to $x=1$. 
Therefore, in the construction of the combined spectrum,
it would make sense to fold the inner continuum twice into the blank of the outer one:
from $x=1$ to $x=0$ and then from $x=0$ to $x=1$.
We let the conversion between the inside/outside coordinates be $\bar{r} = \bar{r}_0 \pm x \Delta \bar{r}  / 2$,
in which $\Delta \bar{r}$ is the full island width in $\bar{r}$ given by the difference
\begin{equation}
    \Delta \bar{r} = \sqrt{\frac{\psi_0 + w/\pi}{\psi_{\text{edge}}}} - \sqrt{\frac{\psi_0 - w/\pi}{\psi_{\text{edge}}}},
\end{equation}
and $\bar{r}_0$ given by the average.
After the conversion, the blank in the $\bar{r}$ coordinate is filled and the continuum is now defined for the entire range $\bar{r} \in [0,1]$.
Such a construction does not have a rigorous mathematical or physical meaning,
but will nevertheless aid us in understanding the width and openness of a frequency gap and its interaction with a global TAE, for instance.
It is noteworthy that the topology of the flux surfaces is different inside/outside the island even if they can be now plotted continuously as a function of $\bar{r}$.
One should avoid the misinterpretation that $\bar{r}$ inside the island labels the effective radius with respect to the magnetic axis as if it is a simple interpolation between the two sides.

\begin{figure}[htbp]
    \centering
    \begin{tabular}{c c}
    \includegraphics[width=7.5cm]{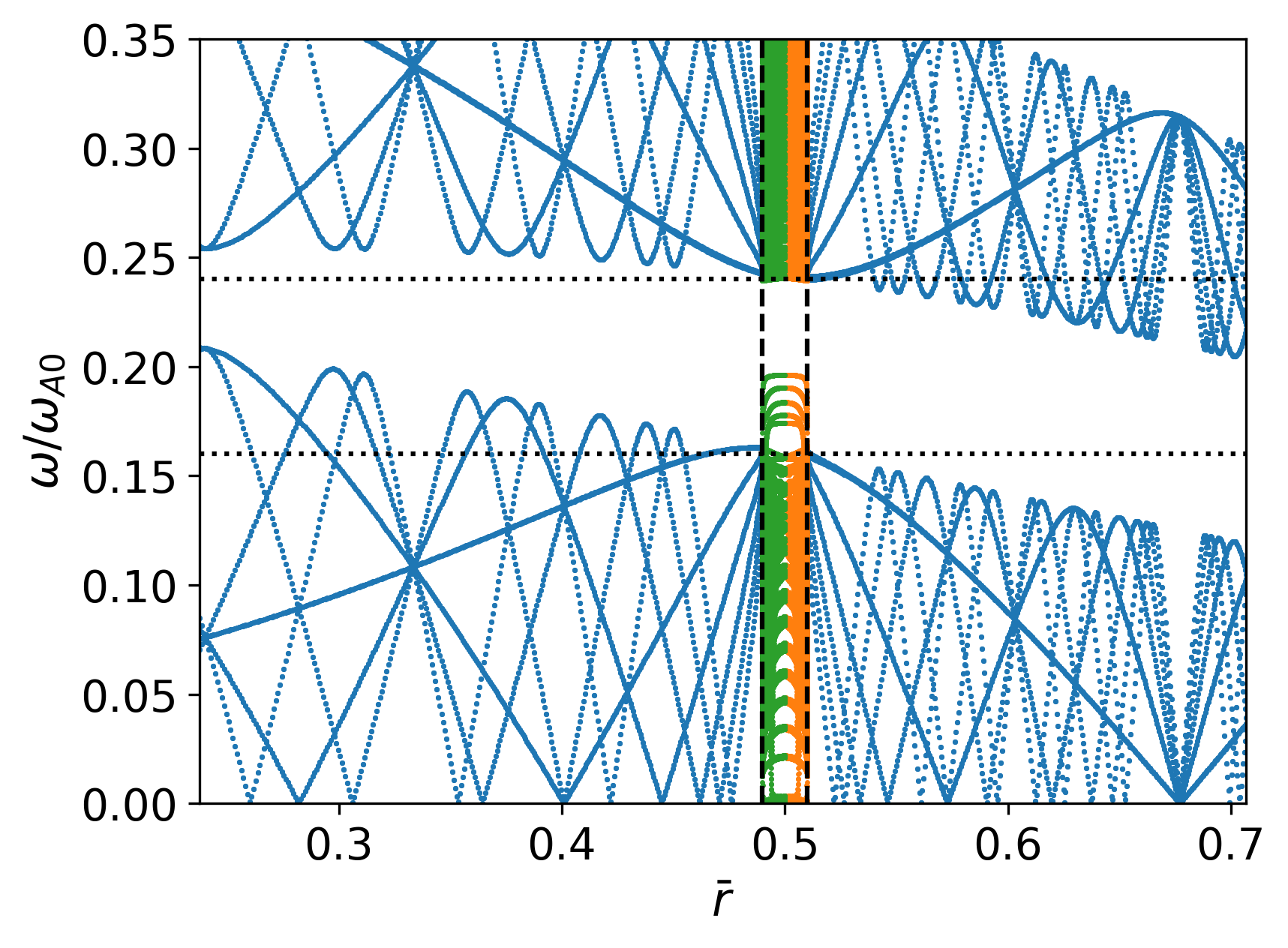} & 
    \includegraphics[width=7.5cm]{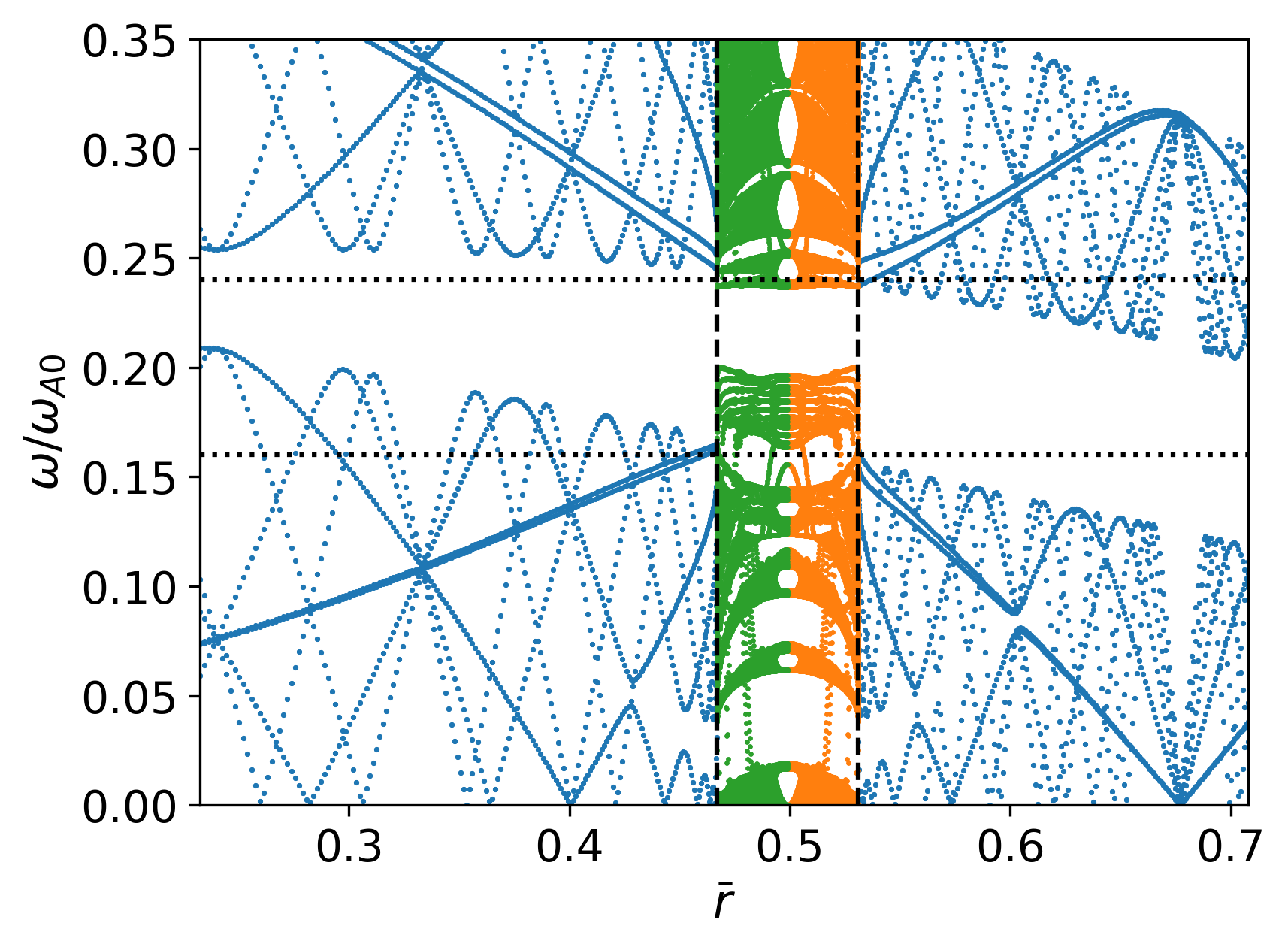}\\
    (a) & (b)   
    \end{tabular}
    \caption{The combined continuum of the $n=1$ family for (a) $A=10^{-5}$ and (b) $A=10^{-4}$.
    The two dotted horizontal lines label the outside TAE gap frequencies in the no-island limit,
    same as \figref{fig:continuum_noisland} (b).
    The vertical dash line indicates the island separatrix. }
    \label{fig:continuum_combined}
\end{figure}

We have plotted the combined continuum spectrum of the $n=1$ family for the two choices of $A$ as shown in \figref{fig:continuum_combined} with an increased number of Fourier modes outside the island.
In both cases, the combined TAE/MAE gap appears much narrower than the TAE gap without an island, indicating a potential intersection with the global mode inside the gap.
Another feature is the continuation of the island EAE gap into the outside region, becoming the HAE gap.
\modi
The same phenomenon was discovered recently by K\"{o}nies \etal~\cite{Konies2022} in a cylindrical plasma and in W7-X,
\norm
who denoted the combined EAE/HAE gap as the global MiAE gap.
All other island gaps are only confined within the island chain.

\section{Conclusions and discussions}\label{sec:conclusions}
In this paper, we have calculated the continuum spectrum for a large aspect ratio tokamak plasma with an island chain.
Two sets of straight-field-line PEST-like coordinates are constructed for inside and outside the island separately, with the shear \Alfven continuum equation being applied to either region.
A Fourier-based numerical scheme was built to find the continuum frequency on flux surfaces both inside and outside, which has been utilised to recover the up-shift of the accumulation point frequency on the island separatrix previously discovered by Biancalani \etal.
We found the outside TAE gap continues into the island region and becomes the MAE gap, 
while the combined TAE/MAE gap has its lower tip shifted up, leading to a much narrower gap compared to the case with no island.
We have also shown that the eigenmode on the lower tip has a localised 2D mode structure, 
accompanied by the fundamental change of the continuous spectrum into a discrete one.
These findings imply that an initially undamped global TAE within the gap may have its frequency intersecting with the continuum inside the island after an island opens up, 
leading to an increased continuum damping.
The estimation of the subsequent damping rate is also complicated by the discrete nature of the modes on the lower tip, which requires further investigation.

\modi
It is noteworthy that the total magnetic field in \eqref{eq:B_components} is not in a self-consistent equilibrium state satisfying force balance,
but rather an axisymmetric equilibrium with an imposed island-opening perturbation.
When a fully self-consistent 3D equilibrium is taken into account, 
it would mean our constant amplitude $A$ becomes a function of the radius,
while multiple Fourier harmonics with different helicities should be included in the magnetic field.
One will also need to add a similar perturbation term to the toroidal flux function $\psi$.
These additional Fourier harmonics would likely induce islands at their own resonance surfaces and interact with each other,
creating a chaotic region around each separatrix and making our current theory invalid.
Luckily, the existence of flux surfaces away from the separatrices (with a sufficiently irrational rotational transform) is guaranteed by the Kolmogorov–Arnold–Moser (KAM) theorem~\cite{Arnold1963}. A small enough perturbation from another Fourier harmonic will slightly displace and deform them but will not affect their topology.
Our present approach avoids treating regions of chaos and is valid when the islands are small with a far enough distance between them, such that they can be considered independently with the chaotic regions being negligibly small.
To construct a fully self-consistent 3D equilibrium with an island chain and study its impact,
one will need a 3D equilibrium code such as HINT~\cite{hayashi1990three}, SIESTA~\cite{Hirshman2011} or SPEC~\cite{Hudson2012, Qu2020}.
This is planned in our future work. 
\norm

\modi
Another notable limitation of the current work is the exclusion of non-ideal effects,
especially the thermal ion finite-Larmor-radius (FLR) effects and the electron parallel dynamics which resolve the ideal MHD continuum as mode conversion layers into the kinetic \Alfven waves~\cite{Hasegawa1975}.
These additional physics are likely important around the separatrix where the inside/outside continuum connects.
Moreover, the fine structures we obtained in the paper, such as the NSAEs, may be modified or replaced by kinetic phenomena when their spatial widths are comparable to or smaller than the ion Larmor radius and/or the electron skin depth. \norm

\modi
The current work can be extended in several other directions.
First, a full parameter scan can be performed to study the dependency of the combined gap width on the island helicity and location.
Second, using the same framework one can compute the continuum for realistic tokamak equilibria in the presence of non-interacting island chains.
The main difference will be to replace our analytical equilibrium with an experimentally relevant one,
with the island described by either an imposed perturbation similar to this work  (same as the CONTI approach) or a nonlinear equilibrium mentioned above.
One will need to construct the straight-field-line coordinates numerically.
Third, we will include the effect of a non-zero pressure profile,
which leads to the up-shift of the CAP frequency and gives rise to the BAE gap.
Also, the global shear \Alfven wave equation \eqref{eq:full_wave} should be solved across the whole plasma to determine the impact of the island continuum on a global TAE.
Furthermore, one could study the effect of island rotation,
which requires a more complicated formalism than the linear ideal MHD equations used in this paper when the island can no longer be considered as a time-independent equilibrium field.
Finally, as an ultimate step, we plan to compute the shear \Alfven continuum and eigenmodes with both islands and chaotic regions.
This will pose interesting physics questions such as the existence of the continuous spectrum in a chaotic region,
since the flux surface on which the mode structure is singular does not exist anymore.
The fine mode structure there may also need the inclusion of non-ideal effects to resolve properly.
\norm

\ack
We thank Stuart Hudson, Nicholas Bohlsen and Robert Dewar for fruitful discussions.
This research was undertaken with the assistance of resources and services from the National Computational Infrastructure (NCI), which is supported by the Australian Government.
This work was supported by a grant from the Simons Foundation/SFARI (560651, AB). 
This work is partly funded by Australian Research Council project DP170102606.

\appendix
\section{Metric tensor and Jacobian in the shifted-circle coordinates}
\label{app:metrics}
To order $O(\epsilon)$, the metrics and the Jacobian are given by
\begin{eqnarray}
    |\gradpsi|^2 = r^2 B_0^2 (1 + 2 \Delta' \cos \vartheta), 
    \label{eq:metric_1} \\
    |\gradt|^2 = \frac{1}{r^2}\left[1 - 2\left(\frac{r}{R_0} + \Delta'\right) \cos\vartheta \right], 
    \label{eq:metric_2} \\
    \gradt \cdot \gradpsi = - B_0 \sin \vartheta \left[ \frac{r}{R_0} + (r \Delta')' \right], 
    \label{eq:metric_3} \\
    |\gradz|^2 = \frac{1}{R_0^2} \left[1 - 2 \left( \frac{r}{R_0} \right) \cos \vartheta \right], 
    \label{eq:metric_4} \\
    J = |\gradpsi \times \gradt \cdot \gradz|^{-1} = \frac{R_0}{B_0} \left(1 + \frac{2r}{R_0} \cos \vartheta \right).
    \label{eq:metric_5} 
\end{eqnarray}
The contra-variant metrics \eqref{eq:metric_1}-\eqref{eq:metric_4} can be inverted to get the co-variant metrics $g_{ij}$ and then the field strength by noting that $B^2 = \sum_{i,j}B^i B^j g_{ij}$.
These quantities are given as a function of $(\psi, \vartheta, \zeta)$.


\section{The action-angle coordinates inside and outside an island chain}
\label{app:island_action_angle}
For outside the island,
\begin{equation}
    \alpha = \frac{2}{m_0} \am  \left( \frac{m_0 K(\kappa^{-1})}{\pi} \baralpha , \kappa^{-1} \right),
\end{equation}
in which $\am(x, \kappa^{-1})$ is the Jacobi amplitude function.
For inside the island,
\begin{equation}
    \alpha = \frac{2}{m_0} \arcsin \left( \sqrt{\kappa} \sn \left(\frac{2 K(\kappa)}{\pi} \baralpha, \kappa \right) \right),
\end{equation}
in which $\sn(x, \kappa)$ is the Jacobi elliptic sin function.

\section{Calculation of the Fourier coefficients}
\label{app:gradpsi}
\begin{eqnarray}
    |\nabla \chi|^2 &=& \frac{q_0'^2}{q_0^4}(\psi-\psi_0)^2 |\gradpsi|^2 + A^2m_0^2 \sin^2 m_0 \alpha 
    \left(|\gradt|^2 + \frac{|\gradz|^2}{q_0^2} \right) \nonumber\\
    &+& 2 \frac{q_0'}{q^2_0}(\psi - \psi_0) A m_0 \sin m_0 \alpha (\gradt \cdot \gradpsi) \nonumber\\
    &+& \frac{q_0'}{q^2_0} w \sqrt{\kappa} A m_0 \cos \baralpha \sin (2\sqrt{\kappa} \sin \baralpha) (\gradt \cdot \gradpsi) \nonumber \\
    &\approx& 4 \Omega_0^2 \kappa \cos^2 \baralpha |\gradpsi|^2
    + 4 \kappa A^2 m_0^2 \sin^2 \baralpha \left(|\gradt|^2 + \frac{|\gradz|^2}{q_0^2} \right) \nonumber\\
    &+& 8 \Omega_0 \kappa A m_0 \cos \baralpha \sin \baralpha (\gradt \cdot \gradpsi). 
 \end{eqnarray}
The last step follows \eqref{eq:theta_reduced} and \eqref{eq:psi_reduced}. Now on the island axis
\begin{eqnarray}
    |\gradpsi|^2 = r_0^2 B_0^2 \left(1 + 2 \Delta' \cos \frac{n_0}{m_0}\zeta \right), \\
    |\gradt|^2 = \frac{1}{r_0^2}\left[1 - 2\left(\epsilon_0 + \Delta'\right) \frac{n_0}{m_0}\zeta \right], \\
    \gradt \cdot \gradpsi = - B_0 \sin \frac{n_0}{m_0}\zeta \left[ \epsilon_0 + (r \Delta')' \right], \\
    |\gradz|^2 = \frac{1}{R_0^2} \left[1 - 2 \epsilon_0 \cos \frac{n_0}{m_0}\zeta \right],
\end{eqnarray}
 all evaluated at $r=r_0$, with $\epsilon_0 = r_0/R_0$.
So to $O(\epsilon)$
\begin{eqnarray}
    |\nabla \chi|^2 &=& 2\Omega_0^2 \kappa r_0^2 B_0^2 (1 + \cos 2 \baralpha)  (1 + 2 \Delta' \cos \frac{n_0}{m_0}\zeta) \nonumber \\
    &+& 2 \kappa \frac{A^2 m_0^2}{r_0^2} (1 - \cos 2 \baralpha) \left[1 - 2\left(\epsilon_0 + \Delta'\right) \cos \frac{n_0}{m_0}\zeta \right]  \nonumber\\
    &-& 4 \Omega_0 \kappa A m_0 B_0 \sin 2 \baralpha  \sin \frac{n_0}{m_0}\zeta \left[ \epsilon_0 + (r \Delta')' \right] \nonumber\\
    &=& 2\Omega_0^2 \kappa r_0^2 B_0^2 \left[ (1 + \delta^2) + (1-\delta^2) \cos^2 2 \baralpha \right.\nonumber\\
    &+&\left.2\left(\Delta' -\delta^2 (\epsilon_0 + \Delta') \right) \cos \frac{n_0}{m_0}\zeta \right.\nonumber\\
    &+& \left. \left( \Delta' + \delta (\epsilon_0 + (r\Delta')') + \delta^2(\epsilon + \Delta') \right)\cos\left(2 \baralpha + \frac{n_0}{m_0}\zeta \right)\right] \nonumber\\
    &+& \left. \left( \Delta' + \delta (\epsilon_0 + (r\Delta')') - \delta^2(\epsilon + \Delta') \right)\cos\left(2 \baralpha - \frac{n_0}{m_0}\zeta \right)\right],
 \end{eqnarray}
 where $\delta = (A m_0)/(\Omega_0 B_0 r_0^2)$.

\section*{References}
\bibliography{references}

\begin{thebibliography}{10}

\bibitem{Fasoli2007}
A.~Fasoli, {\it et~al.\/}, {\it Nuclear Fusion\/} {\bf 47}, S264 (2007).

\bibitem{Heidbrink2008}
W.~W. Heidbrink, {\it Physics of Plasmas\/} {\bf 15}, 1 (2008).

\bibitem{Cheng1985}
C.~Cheng, L.~Chen, M.~Chance, {\it Annals of Physics\/} {\bf 161}, 21 (1985).

\bibitem{Cheng1986}
C.~Z. Cheng, M.~S. Chance, {\it Physics of Fluids\/} {\bf 29}, 3695 (1986).

\bibitem{Furth1973}
H.~P. Furth, P.~H. Rutherford, H.~Selberg, {\it Physics of Fluids\/} {\bf 16},
  1054 (1973).

\bibitem{Evans2006}
T.~E. Evans, {\it et~al.\/}, {\it Nature Physics\/} {\bf 2}, 419 (2006).

\bibitem{Loarte2007}
A.~Loarte, {\it et~al.\/}, {\it Nuclear Fusion\/} {\bf 47}, S203 (2007).

\bibitem{Bortolon2013}
A.~Bortolon, {\it et~al.\/}, {\it Physical Review Letters\/} {\bf 110}, 265008
  (2013).

\bibitem{Kramer2016}
G.~J. Kramer, {\it et~al.\/}, {\it Plasma Physics and Controlled Fusion\/} {\bf
  58}, 085003 (2016).

\bibitem{Kim2020}
K.~Kim, J.~Kang, H.~Kim, J.~Kim, {\it Nuclear Fusion\/} {\bf 60}, 126012
  (2020).

\bibitem{Garcia-Munoz2019}
M.~Garcia-Munoz, {\it et~al.\/}, {\it Plasma Physics and Controlled Fusion\/}
  {\bf 61}, 054007 (2019).

\bibitem{Biancalani2010prl}
A.~Biancalani, L.~Chen, F.~Pegoraro, F.~Zonca, {\it Physical Review Letters\/}
  {\bf 105}, 095002 (2010).

\bibitem{Biancalani2010}
A.~Biancalani, L.~Chen, F.~Pegoraro, F.~Zonca, {\it Physics of Plasmas\/} {\bf
  17}, 122106 (2010).

\bibitem{Biancalani2011}
A.~Biancalani, L.~Chen, F.~Pegoraro, F.~Zonca, {\it Plasma Physics and
  Controlled Fusion\/} {\bf 53}, 025009 (2011).

\bibitem{cook2015shear}
C.~R. Cook, Shear alfv{\'e}n continua and discrete modes in the presence of a
  magnetic island, Ph.D. thesis, The University of Wisconsin-Madison (2015).

\bibitem{Cook2015a}
C.~R. Cook, C.~C. Hegna, {\it Physics of Plasmas\/} {\bf 22}, 042517 (2015).

\bibitem{Yang2022}
J.~Yang, J.~Cao, J.~Zhao, Y.~Dai, D.~Xiang, {\it Plasma Science and
  Technology\/}  (2022).

\bibitem{Dewar1974}
R.~L. Dewar, {\it et~al.\/}, {\it Physics of Fluids\/} {\bf 17}, 930 (1974).

\bibitem{Konies2022}
A.~K{\"{o}}nies, J.~Cao, R.~Kleiber, J.~Geiger, {\it Physics of Plasmas\/} {\bf
  29}, 092102 (2022).

\bibitem{Konies2012}
A.~K{\"{o}}nies, R.~Kleiber, {\it Physics of Plasmas\/} {\bf 19} (2012).

\bibitem{Sun2015}
B.~J. Sun, M.~A. Ochando, D.~L{\'{o}}pez-Bruna, {\it Nuclear Fusion\/} {\bf 55}
  (2015).

\bibitem{Liu2019}
L.~Liu, {\it et~al.\/}, {\it Nuclear Fusion\/} {\bf 59} (2019).

\bibitem{Cook2016}
C.~R. Cook, {\it et~al.\/}, {\it Plasma Physics and Controlled Fusion\/} {\bf
  58}, 054004 (2016).

\bibitem{Hirshman2011}
S.~P. Hirshman, R.~Sanchez, C.~R. Cook, {\it Physics of Plasmas\/} {\bf 18},
  062504 (2011).

\bibitem{Buratti2005}
P.~Buratti, {\it et~al.\/}, {\it Nuclear Fusion\/} {\bf 45}, 1446 (2005).

\bibitem{Annibaldi2007}
S.~V. Annibaldi, F.~Zonca, P.~Buratti, {\it Plasma Physics and Controlled
  Fusion\/} {\bf 49}, 475 (2007).

\bibitem{Chen2011}
W.~Chen, {\it et~al.\/}, {\it Nuclear Fusion\/} {\bf 51} (2011).

\bibitem{Nuhrenberg2000}
C.~N{\"{u}}hrenberg, {\it {ISSP-19 ‘Piero Caldirola', Theory of Fusion
  Plasmas ed JW Connor et al.}\/} ({O Sauter and E Sindoni (Bologna: Editrice
  Compositori–Societa Italiana di Fisica)}, 2000).

\bibitem{Kolesnichenko2001}
Y.~I. Kolesnichenko, V.~V. Lutsenko, H.~Wobig, Y.~V. Yakovenko, O.~P. Fesenyuk,
  {\it Physics of Plasmas\/} {\bf 8}, 491 (2001).

\bibitem{Nakajima1992}
N.~Nakajima, C.~Z. Cheng, M.~Okamoto, {\it Physics of Fluids B: Plasma
  Physics\/} {\bf 4}, 1115 (1992).

\bibitem{Nuhrenberg1999}
C.~N{\"{u}}hrenberg, {\it Plasma Physics and Controlled Fusion\/} {\bf 41},
  1055 (1999).

\bibitem{Salat1992}
A.~Salat, {\it Plasma Physics and Controlled Fusion\/} {\bf 34}, 1339 (1992).

\bibitem{Rutherford1973}
P.~H. Rutherford, {\it Physics of Fluids\/} {\bf 16}, 1903 (1973).

\bibitem{Berk1992}
H.~L. Berk, J.~W. {Van Dam}, Z.~Guo, D.~M. Lindberg, {\it Physics of Fluids B:
  Plasma Physics\/} {\bf 4}, 1806 (1992).

\bibitem{goedbloed2010advanced}
J.~P. Goedbloed, R.~Keppens, S.~Poedts, {\it Advanced magnetohydrodynamics:
  with applications to laboratory and astrophysical plasmas\/} (Cambridge
  University Press, 2010).

\bibitem{Hegna1992}
C.~Hegna, J.~D. Callen, {\it Physics of Fluids B: Plasma Physics\/} {\bf 4},
  3031 (1992).

\bibitem{Hegna2011}
C.~C. Hegna, {\it Nuclear Fusion\/} {\bf 51} (2011).

\bibitem{Rosenbluth1975}
M.~N. Rosenbluth, P.~H. Rutherford, {\it Physical Review Letters\/} {\bf 34},
  1428 (1975).

\bibitem{Spong2003}
D.~A. Spong, R.~Sanchez, A.~Weller, {\it Physics of Plasmas\/} {\bf 10}, 3217
  (2003).

\bibitem{Hirshman1983}
S.~P. Hirshman, J.~C. Whitson, {\it Physics of Fluids\/} {\bf 26}, 3553 (1983).

\bibitem{Kwak2013}
J.~G. Kwak, {\it et~al.\/}, {\it Nuclear Fusion\/} {\bf 53} (2013).

\bibitem{Yakovenko2007}
Y.~V. Yakovenko, {\it et~al.\/}, {\it Plasma Physics and Controlled Fusion\/}
  {\bf 49}, 535 (2007).

\bibitem{Salat1997}
A.~Salat, J.~A. Tataronis, {\it Physics of Plasmas\/} {\bf 4}, 3770 (1997).

\bibitem{Salat2001}
A.~Salat, J.~A. Tataronis, {\it Physics of Plasmas\/} {\bf 8}, 1207 (2001).

\bibitem{Salat2001a}
A.~Salat, J.~A. Tataronis, {\it Physics of Plasmas\/} {\bf 8}, 1200 (2001).

\bibitem{Dinaburg1976}
E.~I. Dinaburg, Y.~G. Sinai, {\it Functional Analysis and Its Applications\/}
  {\bf 9}, 279 (1976).

\bibitem{Sinai1987}
Y.~G. Sinai, {\it Journal of Statistical Physics\/} {\bf 46}, 861 (1987).

\bibitem{Frohlich1990}
J.~Fr{\"{o}}hlich, T.~Spencer, P.~Wittwer, {\it Communications in Mathematical
  Physics\/} {\bf 132}, 5 (1990).

\bibitem{Cuthbert2000}
P.~Cuthbert, R.~L. Dewar, {\it Physics of Plasmas\/} {\bf 7}, 2302 (2000).

\bibitem{Arnold1963}
V.~I. Arnol'd, {\it Russian Mathematical Surveys\/} {\bf 18}, 9 (1963).

\bibitem{hayashi1990three}
T.~Hayashi, T.~Sato, A.~Takei, {\it Physics of Fluids B: Plasma Physics\/} {\bf
  2}, 329 (1990).

\bibitem{Hudson2012}
S.~R. Hudson, {\it et~al.\/}, {\it Physics of Plasmas\/} {\bf 19}, 112502
  (2012).

\bibitem{Qu2020}
Z.~S. Qu, {\it et~al.\/}, {\it Plasma Physics and Controlled Fusion\/} {\bf
  62}, 124004 (2020).

\bibitem{Hasegawa1975}
A.~Hasegawa, L.~Chen, {\it Physical Review Letters\/} {\bf 35}, 370 (1975).

\end{thebibliography}

\end{document}